\documentclass[letterpaper,twocolumn,10pt]{article}

\usepackage{usenix}

\usepackage[utf8]{inputenc}
\usepackage[T1]{fontenc}
\usepackage{microtype}

\usepackage{amsmath,amssymb,amsthm,bm,nicefrac}

\usepackage{tikz}
\usepackage{graphicx}
\usepackage{booktabs}
\usepackage{subcaption}

\usepackage{algorithm}
\usepackage[noend]{algpseudocode}

\usepackage{hyperref} 
\hypersetup{colorlinks=true,
    pdfauthor={Anonymous},
}

\usepackage{url}
\usepackage{multicol}
\usepackage{breqn}
\usepackage{mathtools}
\usepackage{multirow}
\usepackage{xurl}

\newtheorem{lemma}{Lemma}

\newtheorem{definition}{Definition}
\newtheorem{theorem}{Theorem}
\newtheorem{proposition}{Proposition}

\newcounter{footnoteA}

\newcounter{footnoteB}

\begin{document}

\date{}
\title{\Large \bf Locally Private Subgraph Counting via Noisy Adjacency Matrix and Differential Privacy on Randomized Data}

\author{
  {\rm Jintao Guo$^{1}$ (225040008@link.cuhk.edu.cn)} \and
  {\rm Ying Zhou$^{2}$ (22110019@bjtu.edu.cn)} \and
  {\rm GuiXun Luo$^{2}$ (gxluo@bjtu.edu.cn)} \and
  {\rm Chao Li$^{2}$ (li.chao@bjtu.edu.cn)} \and
  {\rm Yan Hu$^{1,3}$ (huyan@cuhk.edu.cn)}
}

\maketitle

\footnotetext[1]{Chinese University of Hong Kong, Shenzhen}
\footnotetext[2]{Beijing Jiaotong University}
\footnotetext[3]{National Health Data Institute, Shenzhen}

\begingroup
\makeatletter
\renewcommand\@makefntext[1]{\noindent #1}
\makeatother
\renewcommand{\thefootnote}{}
\endgroup

\begin{abstract}

Subgraph counting is a fundamental primitive for graph analytics, with applications ranging from social recommendation to anomaly detection. Private subgraph counting under edge local differential privacy (edge-LDP) is challenging without a trusted server or shuffler. We propose two abstractions: the \textbf{Noisy Adjacency Matrix (NAM)} framework, which reformulates private subgraph counting as algebraic estimation over unbiased randomized matrices, and \textbf{Differential Privacy on Randomized Data (DPRD)}, which gives a composition rule for calibrating later-round noise when auxiliary inputs are randomized. Using these tools, we design one-round and two-round estimators for triangle and quadrangle (4-cycle) counting, including TriOR, TriTR, TriMTR, QuaTR, and $\text{TriTR}^*$. We provide the first exact closed-form MSE for the RR-based one-round triangle estimator class and the first relative-error bounds for two-round triangle counting, showing error independent of graph size when the clustering coefficient and average degree are stable. Experiments on two real graphs show that the proposed two-round algorithms outperform all evaluated edge-LDP and shuffle-model baselines in terms of relative error.
\end{abstract}

\begin{figure}[t]
  \centering
  \includegraphics[width=\linewidth]{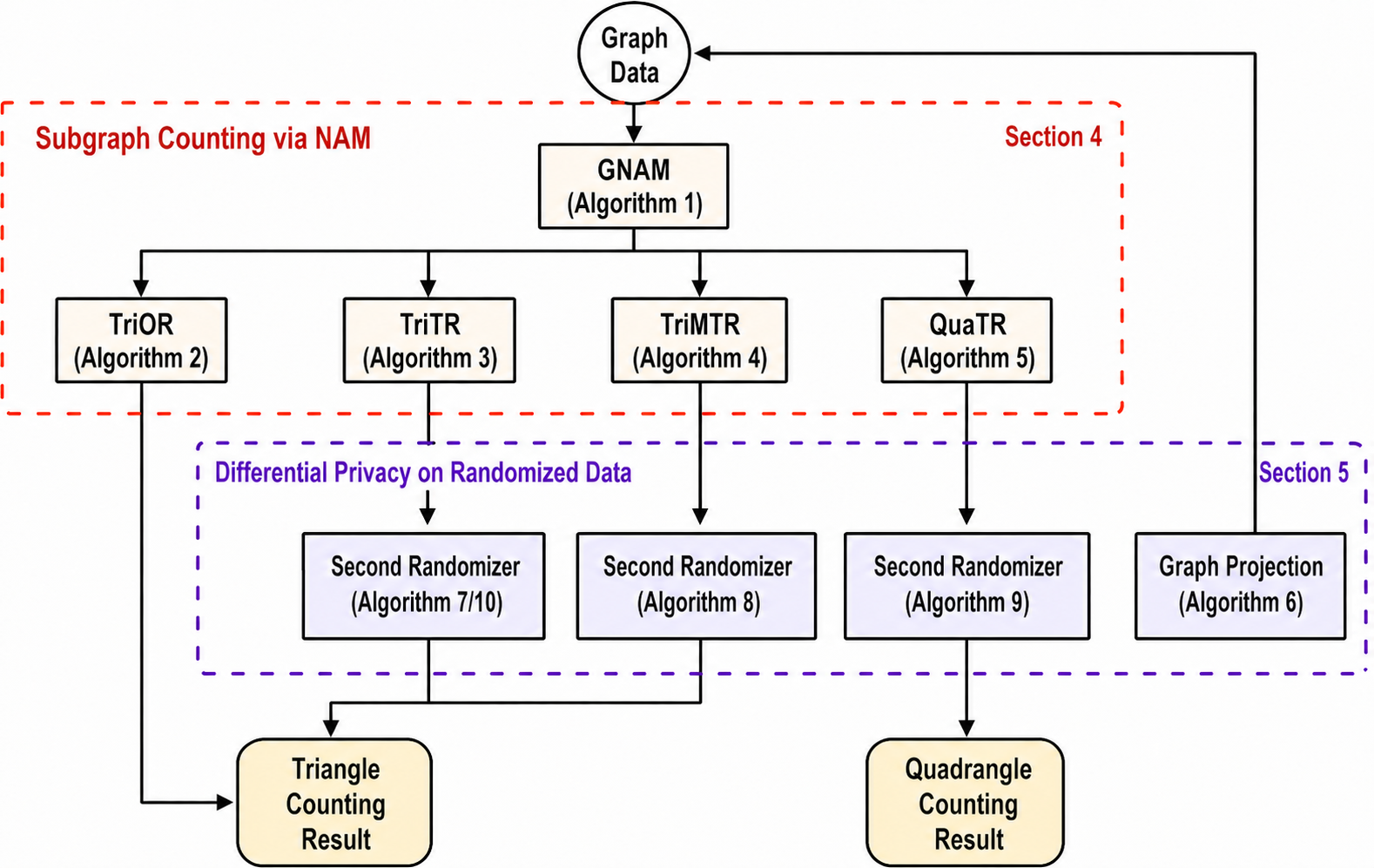}
  \caption{\textbf{Overview of the Dual-Framework Architecture for Private Subgraph Counting.} The workflow is built upon two core pillars: \textbf{NAM} (for algebraic estimation) and \textbf{DPRD} (for interactive privacy security). (1) One-round TriOR operates directly on graph data. (2) Two-round algorithms (TriTR, TriMTR, QuaTR) integrate these frameworks by applying Graph Projection first, then GNAM, and utilizing the Second Randomizer (powered by DPRD) to ensure strict privacy.}
  \label{fig:1}
\end{figure}

\begin{table*}[t]

  \centering

  \caption{\textbf{Theoretical Comparison on Binary Subgraph Counting.} \(\alpha \in [0,1)\), \(C_4\) is the number of cycles of length 4 in \(G\). The expression \(O(n^{2.371552})\) denotes the upper bound on the time complexity of matrix multiplication established in~\cite{williams2024new}. "Small" denotes bias that stems from Graph Projection or Clipping, constrained by their parameters (see~\autoref{sec:5.3}).}

  \resizebox{\textwidth}{!}{

  \begin{tabular}{|c|c|c|c|c|c|c|c|c|}

    \hline

        Subgraph & Algorithm & DP-type & Model & Bias & Variance & \(\text{Time}_{\text{user}}\) & \(\text{Time}_{\text{data collector}}\) & \(\text{Cost}_{DL}\)\\

        \hline

        Triangle & \textbf{TriOR (Ours)} & \(\epsilon\) & one-round & 0 & \(O(nd_{\max}^3+n^3)\) & \(O(n)\) & \({O(n^{2.371552})}\) & 0\\

        Triangle & \(\text{RR}_{\bigtriangleup}\)~\cite{imola2021locally} & \(\epsilon\) & one-round & 0 & \(O(n^4)\) & \(O(n)\) & \(O(n^3)\) & 0\\

        Triangle & \(\text{ARR}_{\bigtriangleup}\)~\cite{imola2022communication} & \(\epsilon\) & one-round & 0 & \(O(n^6)\) & \(O(n)\) & \(O(n^2)\) & 0\\

        Triangle & TCRR~\cite{eden2025triangle} & \(\epsilon\) & one-round & 0 & \({O(C_4+n^3)}\) & \(O(n)\) & \({O(n^{3})}\) & 0\\

        Triangle & \(\text{WLocal}_{\bigtriangleup}\)~\cite{imola2022differentially} & \(\epsilon\) & one-round & 0 & \({O(n^4+n^3d_{\max}^2)}\) & \(O(n)\) & \({O(n^{2})}\) & 0\\

        Quadrangle & \(\text{WLocal}_{\Box}\)~\cite{imola2022differentially} & \(\epsilon\) & one-round & 0 & \({O(n^6+n^2d_{\max}^6)}\) & \(O(n)\) & \({O(n^{2})}\) & 0\\

        \hline
        
        Triangle & \textbf{\(\text{TriTR}\) (Ours)} & \(\epsilon\) & two-round & Small & \(O(nd_{\max}^3)\) & \(O(n+d_{\max}^2)\) & \(O(n^2)\) & \(O(n^2)\)\\

        Triangle & \textbf{TriMTR (Ours)} & \(\epsilon\) & two-round & Small & \({O(nd_{\max}^3+n^2d_{\max})}\) & \(O(n)\) & \(O(n^{2.371552})\) & \(O(n)\)\\

        Quadrangle & \textbf{\(\text{QuaTR}\) (Ours)} & \(\epsilon\) & two-round & Small & \({O(nd_{\max}^5+n^2d_{\max}^3)}\) & \(O(n+d_{\max}^2)\) & \(O(n^{2.371552})\) & \(O(n^2)\)\\

        Triangle & \textbf{\(\text{TriTR}^*\) (Ours)} & \((\epsilon,\delta)\) & two-round & Small & \(O(nd_{\max}^3)\) & \(O(n+d_{\max}^2)\) & \(O(n^2)\) & \(O(n^2)\)\\

        Triangle & \(\text{2R-Large}_{\bigtriangleup}\)~\cite{imola2022communication} & \((\epsilon,\delta)\) & two-round & Small & \(O(nd_{\max}^3)\) & \(O(n+d_{\max}^2)\) & \(O(n^2)\) & \(O(n^2)\)\\

        Triangle & \(\text{2R-Small}_{\bigtriangleup}\)~\cite{imola2022communication} & \((\epsilon,\delta)\) & two-round & Small & \(O(n^2d_{\max}^3)\) & \(O(n)\) & \(O(n^2)\) & \(O(n \log n)\)\\

        Triangle & TDP-PC~\cite{he2025robust} & \(\epsilon\) & two-round & 0 & \(O(nd_{\max}^3+n^2)\) & \(O(nd_{\max})\) & \(O(n^2)\) & \(O(n^2)\)\\
        
        Triangle & TDP-VC~\cite{he2025robust} & \(\epsilon\) & two-round & Small & \(O(nd_{\max}^3)\) & \(O(n+d_{\max}^2)\) & \(O(n^2)\) & \(O(n^2)\)\\

        \hline

        Triangle & \(\text{WShuffle}_{\bigtriangleup}\)~\cite{imola2022differentially} & \((\epsilon,\delta)\) & shuffle & 0 & \(O(n^3d_{\max}^2)\) & \(O(n)\) & \(O(n^2)\) & 0\\

        Triangle & \(\text{WShuffle}_{\bigtriangleup}^*\)~\cite{imola2022differentially} & \((\epsilon,\delta)\) & shuffle & \(n d_{\text{avg}}^2\) & \(O(n^2d_{\max}^4 + n^{1+2\alpha} d_{\max}^2 )\) & \(O(n)\) & \(O(n^2)\) & 0\\

        Quadrangle & \(\text{WShuffle}_{\Box}\)~\cite{imola2022differentially} & \((\epsilon,\delta)\) & shuffle & 0 & \(O(n^3d_{\max}^2+n^2d_{\max}^6)\) & \(O(n)\) & \(O(n^2)\) & 0\\

        Triangle & \(\text{GenShuffle}_{\bigtriangleup}\)~\cite{fang2025counting} & \((\epsilon,\delta)\) & shuffle & 0 & \(O(n^2d_{\max}+nd_{\max}^{3})\) & \(O(n)\) & \(O(n^2+nd_{\max}^{1.5})\) & 0\\

        Quadrangle & \(\text{GenShuffle}_{\Box}\)~\cite{fang2025counting} & \((\epsilon,\delta)\) & shuffle & 0 & \(O(n^2d_{\max}^2+nd_{\max}^{4.5})\) & \(O(n)\) & \(O(n^2+nd_{\max}^{1.5})\) & 0\\

        \hline

  \end{tabular}

  }

  \label{tab:1}

\end{table*}

\section{Introduction}

Subgraph counting is a fundamental primitive for analyzing graph connectivity, with applications ranging from social recommendation to anomaly detection. In decentralized settings where no trusted server exists, \textit{Edge Local Differential Privacy (Edge-LDP)}~\cite{qin2017generating,imola2021locally,imola2022communication,imola2022differentially,eden2025triangle, he2025robust} has become a standard privacy model for protecting users' sensitive relationships while estimating graph statistics. Beyond binary graphs, weighted subgraph analytics, e.g., estimating the intensity of closed flow cycles in financial networks, raise similarly important privacy concerns and remain largely unexplored under edge-LDP.

Existing edge-LDP algorithms face a stringent privacy--utility--communication trade-off. Non-interactive protocols~\cite{imola2021locally,imola2022communication,imola2022differentially,eden2025triangle} are unbiased but suffer from large variance, and existing analyses provide only loose asymptotic MSE bounds. Two-round interactive protocols~\cite{imola2022communication,he2025robust} significantly improve accuracy, yet they force a sharp choice: users either download a large noisy graph, incurring \(O(n^2)\) communication, or accept substantially degraded accuracy under low download cost. Shuffle-model algorithms~\cite{imola2022differentially, fang2025counting} achieve strong utility, but their guarantees rely on a semi-trusted shuffler; if the shuffler colludes with the data collector, the protection level may degrade considerably~\cite{cheu2019distributed, erlingsson2019amplification}. Given these vulnerabilities, in this work we adopt a strict \textit{threat model}: no trusted server or shuffler exists, the data collector may be malicious, each user holds only their own adjacency list and randomizes their data locally before sending, and any auxiliary information released during multi-round interaction (e.g., noisy adjacency matrices or noisy degrees) is treated as public DP-protected output.

To address these issues, we propose a dual-framework architecture built upon two conceptual pillars, as illustrated in \autoref{fig:1}. First, we introduce the \textbf{Noisy Adjacency Matrix (NAM)} framework, which reformulates private subgraph counting as an algebraic estimation problem on randomized matrices. Second, we introduce the \textbf{Differential Privacy on Randomized Data (DPRD)} framework, which rigorously resolves the privacy challenge arising when a later-round input depends on previous randomized outputs. These two frameworks together enable accurate multi-round subgraph counting under strict edge-LDP without relying on trusted third parties.

\noindent \textbf{Our Contributions.} Our main contributions are as follows:

\begin{itemize}
    \item \textbf{The NAM Framework.} We propose a unified algebraic estimation framework for private subgraph counting. It reveals that the MSE of the proposed algorithms can be compactly expressed via Frobenius norms of powers of the adjacency matrix, and it naturally accommodates different local randomizers as well as binary, weighted, and directed graph extensions.

    \item \textbf{The DPRD Framework.} We formalize the problem of ensuring privacy when the input to a randomized mechanism is itself randomized data, and propose the DPRD framework to handle it. DPRD composes with standard DP and provides a general basis for rigorous privacy proofs in interactive protocols without trusted third parties.

    \item \textbf{Theoretical and Algorithmic Results.} We prove that TriOR is mathematically equivalent to existing one-round triangle algorithms and provide the \textbf{first exact closed-form MSE} for this class. Building on the two frameworks, we obtain several new algorithms: \(\text{TriTR}^*\) reduces second-round noise compared with TriTR/TDP-VC; TriMTR attains the best accuracy among two-round algorithms with \(O(n)\) per-user download cost under our evaluation; and QuaTR is a pure-LDP quadrangle algorithm that outperforms the evaluated shuffle-model baselines. We give relative-error bounds for two-round algorithms, showing error independent of graph size when clustering coefficient and average degree are stable.
\end{itemize}

\section{Related Work}
\label{sec:related_work}

\textbf{Private Subgraph Counting.} 
Our work considers edge-LDP~\cite{ye2020towards, qin2017generating, ye2020lf, sun2019analyzing}, which protects edges without a trusted server. This setting differs from Central DP, which relies on a trusted server~\cite{ding2021differentially, karwa2011private, zhang2015private}, and Node DP, which often introduces excessive noise for graph statistics~\cite{kasiviswanathan2013analyzing, hay2009accurate, day2016publishing, zhang2020differentially, raskhodnikova2016differentially}. \autoref{tab:1} summarizes existing edge-LDP and shuffle-based subgraph counting algorithms alongside ours, categorized into pure LDP (one-round and two-round) and the shuffle model.

\textbf{One-Round Algorithms.} 
Non-interactive subgraph counting typically suffers from low accuracy, and the theoretical MSE upper bounds provided by existing works are loose. We show that the randomization mechanisms of \(\text{RR}_{\bigtriangleup}\) and TCRR are mathematically equivalent to that of our TriOR. While prior work only gives asymptotic bounds, we provide the first \textbf{exact closed-form expression} for the MSE of this algorithm class (\autoref{TriOR}). This analysis confirms their theoretical optimality, as they attain the \(\Omega(n^2)\) lower bound established by Eden et al.~\cite{eden2025triangle}.

\textbf{Two-Round Algorithms.} 
To reduce estimation variance, recent works introduce interactive protocols. The \(\text{2R-Large}_{\triangle}\) and \(\text{2R-Small}_{\triangle}\) are instantiations of the \(\text{ARROneNS}_{\triangle}\) algorithm~\cite{imola2022communication} with parameters \(\mu = \frac{e^{\epsilon_1}}{e^{\epsilon_1}+1}\) and \(\mu = \frac{1}{\sqrt{n}} \frac{e^{\epsilon_1}}{e^{\epsilon_1}+1}\), respectively. Yizhang He et al.~\cite{he2025robust} observe that the algorithm in~\cite{imola2022communication} actually preserves \((\epsilon, \delta)\)-edge LDP, rather than the \(\epsilon\)-edge LDP claimed in~\cite{imola2022communication}, and they propose two improved algorithms, TDP-PC and TDP-VC. Our TriTR is mathematically equivalent to the current state-of-the-art algorithm TDP-VC. To improve upon this, we design \(\text{TriTR}^*\), which further reduces the noise introduced in the second round and thus achieves higher accuracy. Moreover, compared with \(\text{2R-Small}_{\triangle}\), our TriMTR significantly improves accuracy under low download cost.

\textbf{Shuffle Model Algorithms.} 
The shuffle model~\cite{imola2022differentially, fang2025counting, cheu2019distributed, erlingsson2019amplification} enables accurate subgraph counting via an intermediate shuffler. However, its security relies on a strong assumption that the shuffler does not collude with the data collector. If this assumption fails, the \(2\epsilon\)-edge DP guarantee degrades to the \(\epsilon_L\)-edge LDP level, typically with \(\epsilon_L \gg \epsilon\). Our work achieves comparable utility under strict \(2\epsilon\)-edge DP and \(\epsilon\)-edge LDP without relying on semi-trusted third parties. Furthermore, our QuaTR also surpasses these shuffle algorithms in accuracy.

No single algorithm outperforms all others in every scenario; the best choice varies with the application. We summarize their respective application scenarios in \autoref{app:deployment-guidance} to help users select the most suitable one for deployment.

\section{Preliminaries}

\subsection{Notations}
Let $\mathbb{R}$ and $\mathbb{R}_{\ge 0}$ denote the sets of real and non-negative real numbers, respectively. 
Let $\mathbb{N}$ and $\mathbb{Z}_{\ge 0}$ denote the sets of natural numbers and non-negative integers, and let $\emptyset$ denote the empty set. 
For $n \in \mathbb{N}$, define $[n] \triangleq \{1,2,\ldots,n\}$.

Let $\mathcal{G}$ denote the set of all undirected simple graphs (i.e., without self-loops). 
For $G \in \mathcal{G}$, write $G=(V,E)$, where $V=\{v_1,\ldots,v_n\}$ is the node set and $E \subseteq \{\{v_i,v_j\}: i \ne j\}$ is the edge set. 
Let $n \triangleq |V|$ and $m \triangleq |E|$. 
For labeled nodes, let $d_i$ denote the degree of $v_i$, and define $d_{\max} \triangleq \max_{i\in[n]} d_i$ and $d_{\mathrm{avg}} \triangleq \frac{1}{n}\sum_{i=1}^n d_i$.

Let $A=(a_{ij}) \in \{0,1\}^{n\times n}$ denote the adjacency matrix of $G$, where $a_{ij}=1$ iff $\{v_i,v_j\}\in E$ and $a_{ij}=0$ otherwise. 
Since $G$ is undirected and loop-free, $A$ is symmetric and $a_{ii}=0$ for all $i$. 
Let $\mathbf{a}_i \in \{0,1\}^n$ denote the $i$-th row of $A$ (the adjacency list of $v_i$). 
Define $B \triangleq A^2=(b_{ij})$ and $C \triangleq A^3=(c_{ij})$, which we refer to as the two-step and three-step walk matrices, respectively.

Let $f^{\triangle}:\mathcal{G}\to\mathbb{Z}_{\ge 0}$ and $f^{\Box}:\mathcal{G}\to\mathbb{Z}_{\ge 0}$ denote the triangle- and quadrangle-counting functions. 
Given $G\in\mathcal{G}$, they output the number of triangles $f^{\triangle}(G)$ and the number of quadrangles $f^{\Box}(G)$, respectively.

\subsection{Differential Privacy}

\textbf{Definition of Differential Privacy.} Differential privacy (DP) was first proposed by Dwork et al.~\cite{dwork2006calibrating}, and they formalized its theoretical framework in~\cite{dwork2014algorithmic}.

\begin{definition}[\((\epsilon, \delta)\)-Differential Privacy~\cite{dwork2014algorithmic}]\label{def:1}
Let \( \mathcal{M}: \mathcal{X}^n \rightarrow \mathcal{Y} \) be a randomized mechanism, where \( \mathcal{X}^n \) is the space of datasets containing \( n \) data points, and \( \mathcal{Y} \) is the output space. For any two neighboring datasets \( D \) and \( D' \) (i.e., \( D \) and \( D' \) differ in only one record), and for any subset of outputs \( S \subseteq \mathcal{Y} \), if the mechanism \( \mathcal{M} \) satisfies:
\begin{equation}
\Pr[\mathcal{M}(D) \in S] \leq e^\epsilon \cdot \Pr[\mathcal{M}(D') \in S] + \delta
\end{equation}
then the mechanism \( \mathcal{M} \) is said to satisfy \( (\epsilon, \delta) \)-differential privacy.
\end{definition}

If \(\delta = 0\), it is commonly abbreviated as \(\epsilon\)-DP. \(\epsilon\) is typically referred to as the privacy budget, a smaller \(\epsilon\) indicates stronger privacy protection.

\textbf{Implementing Differential Privacy.} Differential privacy can be instantiated via a variety of mechanisms~\cite{dwork2014algorithmic,mcsherry2009privacy,wang2016using}. In this work, we primarily use two widely adopted mechanisms—\emph{randomized response} (RR) and the \emph{Laplace mechanism}—to implement DP.

\textbf{Laplace Mechanism~\cite{dwork2014algorithmic}.} Let \( f: \mathcal{X}^n \rightarrow \mathbb{R}^k \) be a function that maps a dataset \( D \in \mathcal{X}^n \) to a vector of real numbers. The \textit{global sensitivity} of \( f \), denoted by \( \Delta f \), is defined as:

\begin{equation}
\Delta f = \max_{D, D'} \| f(D) - f(D') \|_1
\label{eq:2}
\end{equation}

where the maximum is taken over all pairs of neighboring datasets \( D \) and \( D' \) that differ in at most one record. For a given privacy parameter \( \epsilon > 0 \), the Laplace Mechanism releases:

\begin{equation}
\mathcal{M}_L(D, f, \epsilon) = f(D) + (Y_1, Y_2, \dots, Y_k)
\label{eq:3}
\end{equation}

where \( Y_1, Y_2, \dots, Y_k \) are independent and identically distributed (i.i.d.) random variables drawn from the Laplace distribution \( \text{Lap}\left(\frac{\Delta f}{\epsilon}\right) \).

\textbf{Warner's RR~\cite{warner1965randomized}.} Given \(\epsilon \in \mathbb{R}_{\geq 0}\), \(\mathcal{R} _{\epsilon}^{W}: \left\{ 0,1 \right\} \rightarrow \left\{ 0,1 \right\} \) maps \(x\in \left\{ 0,1 \right\} \) to \(y\in \left\{ 0,1 \right\} \) with the probability:

\begin{equation}
\label{eq:4}
\Pr \left[ \mathcal{R} _{\epsilon}^{W}\left( x \right) =y \right] =\begin{cases}
	\frac{e^{\epsilon}}{e^{\epsilon}+1}\,\,  \left( \text{if } x=y \right)\\
	\frac{1}{e^{\epsilon}+1}\,\,  \left( \mathrm{otherwise} \right)\\
\end{cases}
\end{equation}

\subsection{Differential Privacy on Graphs}

We use the definition of (\(\epsilon, \delta\))-edge DP and (\(\epsilon, \delta\))-edge LDP as our privacy metric. Edge DP in the central model~\cite{raskhodnikova2016differentially} considers two graphs that differ in one edge. In contrast, edge LDP~\cite{qin2017generating} considers two neighbor lists that differ in one bit. 

\begin{definition}[(\(\epsilon, \delta\))-edge DP~\cite{raskhodnikova2016differentially}.] 
\label{def:2}
Let $n \in \mathbb{N}$, $\epsilon \in \mathbb{R}_{\geq 0}$, and $\delta \in [0,1]$. A randomized algorithm $\mathcal{M}$ with domain $\mathcal{G}$ provides ($\epsilon$,$\delta$)-edge DP if for any two neighboring graphs $G, G' \in \mathcal{G}$ that differ in one edge and any $S \subseteq \mathrm{Range}(\mathcal{M})$,
\begin{equation}
\Pr[\mathcal{M}(G) \in S] \leq e^{\epsilon} \Pr[\mathcal{M}(G') \in S] + \delta.
\end{equation}
\end{definition}

\begin{definition}[(\(\epsilon, \delta\))-edge LDP~\cite{qin2017generating}.] 
\label{def:3}
Let \(\epsilon \in \mathbb{R}_{\geq 0}\), and $\delta \in [0,1]$. A local randomizer \(\mathcal{R}\) with domain \(\{0, 1\}^n\) provides (\(\epsilon, \delta\))-edge LDP if for any two neighbor lists \(\mathbf{a}_i, \mathbf{a}_i' \in \{0, 1\}^n\) that differ in one bit and any \(S \subseteq \text{Range}(\mathcal{R})\),
\begin{equation}
\Pr[\mathcal{R}(\mathbf{a}_i) \in S] \leq e^\epsilon \Pr[\mathcal{R}(\mathbf{a}_i') \in S]+ \delta.
\end{equation}
\end{definition}

If \(\delta = 0\), they are generally abbreviated as \(\epsilon\)-edge DP and \(\epsilon\)-edge LDP. Although the definitions of edge DP and edge LDP differ, edge LDP can transfer to edge DP through \autoref{pro:1}.

\begin{proposition}[Edge DP and Edge LDP~\cite{imola2022differentially}]
\label{pro:1}
If a randomized algorithm $\mathcal{M}$ in the central model applies a local randomizer $\mathcal{R}$ providing $\epsilon$-edge LDP to each neighbor list $\mathbf{a}_{i}$ ($1 \leq i \leq n$), it provides $2\epsilon$-edge DP.
\end{proposition}

\textbf{Interaction among Users and Multiple Rounds.} It is common to provide users with auxiliary information through multiple rounds of queries to generate a more accurate estimate. Therefore, we leverage the sequential composition of edge LDP to ensure privacy guarantees.

\begin{proposition}[Sequential Composition of Edge LDP~\cite{imola2022communication}]
For \( i \in [n] \), let \( \mathcal{R}_{i}^{1} \) be a local randomizer of user \( v_i \) that takes \( \mathbf{a}_{i} \) as input. Let \( \lambda_{i} \) be a post-processing algorithm on \( \mathcal{R}_{i}^{1}(\mathbf{a}_{i}) \) and \( M_{i} = \lambda_{i}(\mathcal{R}_{i}^{1}(\mathbf{a}_{i})) \) be its output. Let \( \mathcal{R}_{i}^{2}(M_{i}) \) be a local randomizer of \( v_{i} \) that depends on \( M_{i} \). If \( \mathcal{R}_{i}^{1} \) provides \( \epsilon_{1} \)-edge LDP and for any \( M_{i} \in \operatorname{Range}(\lambda_{i}) \), \( \mathcal{R}_{i}^{2}(M_{i}) \) provides \( \epsilon_{2} \)-edge LDP, then the sequential composition \( (\mathcal{R}_{i}^{1}(\mathbf{a}_{i}), \mathcal{R}_{i}^{2}(M_{i})(\mathbf{a}_{i})) \) provides \( (\epsilon_{1} + \epsilon_{2}) \)-edge LDP.
\label{pro:2}
\end{proposition}

Due to space constraints, we give the definition of DP for weighted graphs in \autoref{app:weighted_dp}.

\subsection{Utility Metrics}

We evaluated the utility of the algorithm from two perspectives: accuracy and cost of data transmission.

\textbf{Accuracy.} We evaluate accuracy using two metrics: mean squared error (MSE) and relative error (RE). For an estimator $\hat{\theta}$ of a target quantity $\theta$, the MSE is $\mathrm{MSE}(\hat{\theta}) \triangleq \mathbb{E}\!\left[(\hat{\theta}-\theta)^2\right]$,
and the RE is $\mathrm{RE}(\hat{\theta}) \triangleq \mathbb{E}\!\left[\frac{|\hat{\theta}-\theta|}{\theta}\right]$. MSE is convenient for theoretical analysis, whereas RE is often more indicative of practical performance in real applications.

\textbf{Data Transmission Cost.} We measure communication overhead in terms of (i) the per-user upload cost and (ii) the per-user download cost. For most existing edge-LDP subgraph counting protocols, the upload cost per user is at most $O(n)$ bits, and is usually not the dominant bottleneck in practice. Instead, the main limitation comes from the potentially large download volume incurred across multiple interactive rounds, which we denote by $\mathrm{Cost}_{\mathrm{DL}}$. Consider $n$ users and an $r$-round protocol. Let $M_i^{(j)}$ denote the download volume (in bits) received by user $i$ in round $j\in[r]$. We define the total download cost as

\begin{equation}
\text{Cost}_{DL} = \max_{i=1}^{n} \sum_{j=1}^{r} \mathbb{E}[|M_{i}^{j}|] \quad \text{(bits)}.
\end{equation}

\section{Subgraph Counting Algorithms}
\label{sec:4}

In this section, we first define the \textbf{Noisy Adjacency Matrix (NAM)} and analyze its properties. We then develop four algorithms for triangle and quadrangle counting. Finally, we summarize their relationships from a matrix perspective and discuss how to extend the framework to directed graphs.

For two-round algorithms, \autoref{alg:3}--\ref{alg:5} show the idealized second-round computations; their privacy-preserving instantiations are specified in \autoref{sec:5.4} via DPRD.

\subsection{Noisy Adjacency Matrix}
\label{sec:4.1} 

\textbf{Definition and Property.} We define the NAM in a general real-valued form. Let \(A=(a_{ij})\in\mathbb{R}^{n\times n}\) be an adjacency matrix with \(a_{ii}=0\) for any \(i\in[n]\). Here \(a_{ij}\) can be either a binary edge indicator or a real-valued edge weight. Therefore, binary graphs are a special case where \(A\in\{0,1\}^{n\times n}\).

\begin{definition}[Noisy adjacency matrix] 
\label{def:4}
    \(\hat{A}\) is the noisy adjacency matrix of \(A\), if \(\hat{A}\) satisfies:
    \begin{center}
    \(\mathbb{E} \left[ \hat{A} \right] = A; \text{ }\hat{a}_{ii}=0, \text{ for any } i \in [n];\)

    \(\hat{a}_{ij}\bot \hat{a}_{kl}, \text{ for any distinct edge coordinates } (i,j)\ne(k,l).\)
    \end{center}
    For undirected graphs, we additionally require \(\hat{A}=\hat{A}^\top\), and the independence condition is imposed on \(\hat{a}_{ij}\) for \(i<j\).
\end{definition}

For a real-valued adjacency matrix \(A \in \mathbb{R}^{n \times n}\), the \(k\)-th power of \(A\) has the following property:

\begin{proposition}
\label{pro:3}
The value of the element in the \(i\)-th row and \(j\)-th column of \(A^k\) equals the total weight of all length-\(k\) walks from node \(v_i\) to node \(v_j\), where the weight of a walk is the product of the weights of its edges. In the binary case, this value reduces to the number of paths for node \(v_i\) to reach node \(v_j\) exactly after \(k\) steps.
\end{proposition}

\autoref{pro:3} establishes the relationship between walks in a graph and powers of its adjacency matrix. Through our research, we have discovered that the NAM exhibits similar properties.

\begin{theorem}
\label{thm:1}
Let \( A \) and \( \hat{A} \) be the adjacency matrix and the noisy adjacency matrix, respectively. Then:
\begin{enumerate}
    \item Let \(\hat{B}=\hat{A}^2\), \(B=A^2\). \(\mathbb{E} \left[ \hat{b}_{ij} \right] =b_{ij}\), for any \(i\ne j\).
    \item Let \(\hat{C}=\hat{A}^3\), \(C=A^3\). \(\mathbb{E} \left[ \hat{c}_{ii} \right] =c_{ii}\), for any \(i \in [n]\).
\end{enumerate}
\end{theorem}

It is worth noticing that \autoref{thm:1} holds for \(A\in\mathbb{R}^{n\times n}\) and does not rely on the binary nature of adjacency entries. However, since existing works~\cite{imola2021locally, imola2022communication, imola2022differentially, eden2025triangle, he2025robust, fang2025counting} mainly focus on subgraph counting on binary undirected graphs, the main text of this paper also concentrates on this setting. We place the application of NAM to weighted and directed graphs in \autoref{app:weighted}; their algorithmic versions can be obtained by minor adjustments to the main algorithms, as they share the NAM framework and the subsequent DPRD framework. Next, we will show how to integrate the NAM with edge LDP, which will serve as the basis for all subsequent algorithms.

\begin{algorithm}[H]
\caption{GNAM}
\label{alg:1}
\begin{algorithmic}[1]
\Statex \textbf{Input:} Privacy budget \(\epsilon \in \mathbb{R}_{\geq 0}\), user's adjacency lists \(\mathbf{a}_1,\mathbf{a}_2,\ldots,\mathbf{a}_n\).
\Statex \textbf{Output:} Noisy adjacency matrix \(\hat{A}\).
\For{\textbf{each} \textit{User} \(u=1\) to \(n\)}\textbf{:}
    \State \(\tilde{\mathbf{a}}_u \gets \textit{local randomizer}\left( \mathbf{a}_u, \epsilon \right)\)
    \State \(\tilde{a}_{ui} \gets 0, \textit{ for any } i \geqslant u.\)
    \State \textbf{send} \(\tilde{\mathbf{a}}_u\) to \textit{Data Collector}
\EndFor
\Statex \textit{Data Collector} \textbf{do:}
\State \(L \gets (\tilde{\mathbf{a}}_1,\ldots,\tilde{\mathbf{a}}_n)\)
\State \(\tilde{A} \gets L+L^\top\)
\State \( \hat{A} \gets \textit{estimate algorithm}\left(\tilde{A}\right) \)
\State \textbf{return} \(\hat{A}\)
\end{algorithmic}
\end{algorithm}

\textbf{Generating NAM.} We design the Generate Noisy Adjacency Matrix (GNAM) in \autoref{alg:1}. Each user \(u\) applies a \emph{local randomizer} to privatize its adjacency row \(\mathbf{a}_u\), producing a noisy vector \(\tilde{\mathbf{a}}_u\). To avoid double-reporting edges in undirected graphs, user \(u\) sets entries corresponding to higher-index nodes to zero and sends \(\tilde{\mathbf{a}}_u\) to the data collector. The data collector stacks the received vectors to form a lower-triangular noisy matrix \(L\), and then completes the other half using symmetry. Finally, it obtains unbiased estimators for each element, resulting in the noisy adjacency matrix \(\hat{A}\). GNAM holds the LDP guarantee:

\begin{proposition}
\label{pro:GNAM}
GNAM preserves \(\epsilon\)-edge LDP and \(2\epsilon\)-edge DP in the binary undirected graph setting.
\end{proposition}

\begin{figure}[h]
  \centering
  \includegraphics[width=\linewidth]{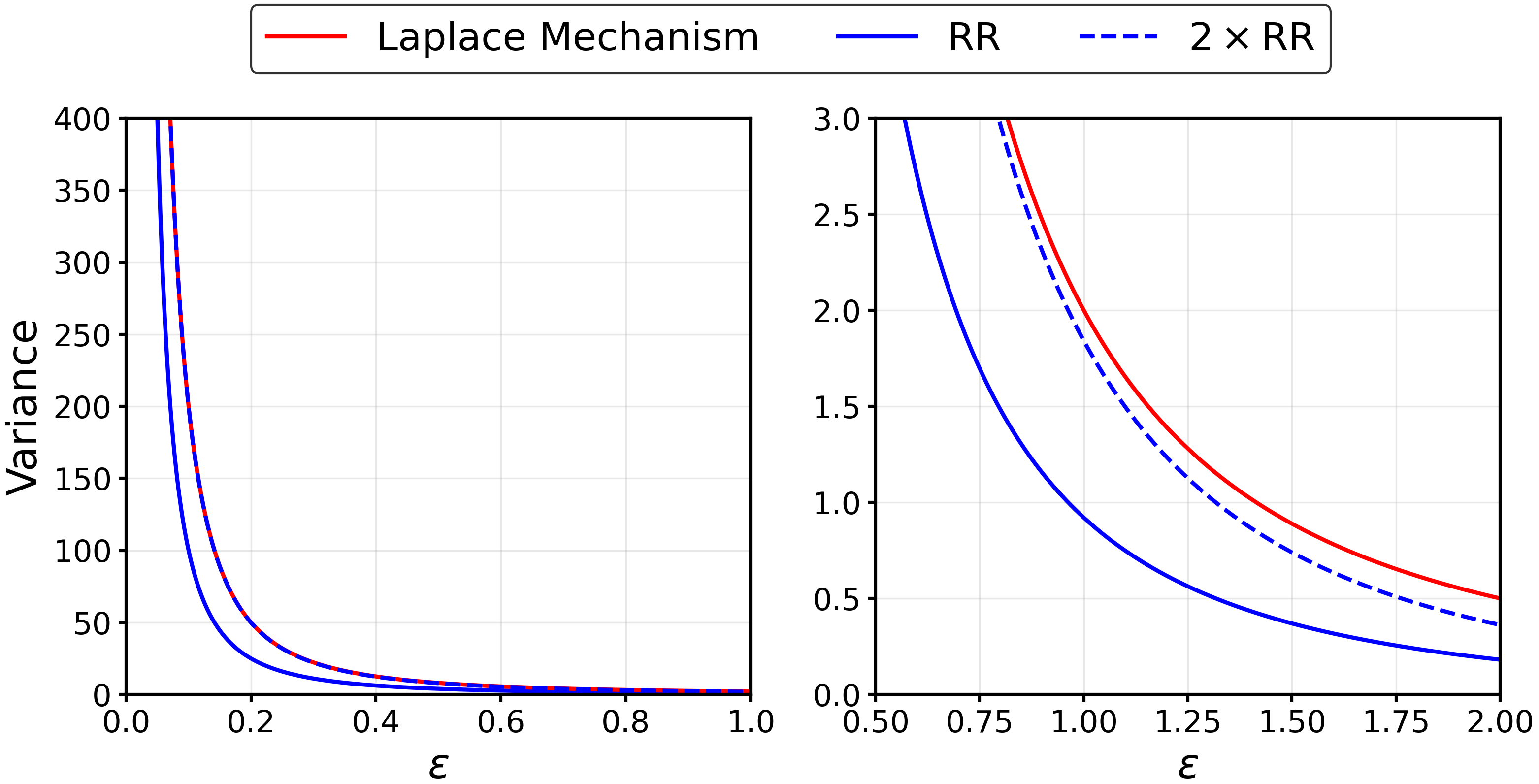}
  \caption{\textbf{Variance Comparison between RR and Laplace Mechanism in Noisy Adjacency Matrix.}  The red line denotes the Laplace mechanism, while the solid blue line represents RR. The dashed blue line indicates $2 \times$ the variance of RR. The comparison highlights the variance trends across varying privacy budgets ($\epsilon$).}
  \label{fig:2}
\end{figure}

The \textit{local randomizer} in Line 2 can implement the RR or Laplace mechanisms. If RR is chosen, Line 2 randomizes each element in \(\mathbf{a}_u\) according to \autoref{eq:4}. Then, Line 7 can apply \autoref{pro:4} to obtain unbiased estimators for each matrix element. If the Laplace mechanism is chosen, Line 2 adds independent Laplace noise to each element of \(\mathbf{a}_u\). Since the Laplace mechanism applies to real-valued inputs, it provides a natural way to generate NAM for weighted graphs. Previous works~\cite{imola2021locally, imola2022communication, imola2022differentially, eden2025triangle, he2025robust, fang2025counting} employ RR to produce noisy matrices on binary undirected graphs. Motivated by these ideas, we generalize the approach to weighted graphs and to other DP mechanisms, while the core framework remains adaptable.

\begin{proposition}
\label{pro:4}
Let \(X\) be the output of Warner's RR, i.e., \(X=\mathcal{R}_{\epsilon}^{W}(a)\), where \(a\in\{0,1\}\). Then \(Y=\frac{X\cdot(e^\epsilon+1)-1}{e^\epsilon-1}\) satisfies \(\mathbb{E}[Y]=a\).
\end{proposition}

The choice between the RR mechanism and the Laplace mechanism leads to different variances for the entries of \(\hat{A}\). Their variances are \(\sigma^2 = \frac{e^{\epsilon}}{(e^{\epsilon}-1)^2}\) and \(\sigma^2 = \frac{2}{\epsilon^2}\), respectively. \autoref{fig:2} shows the relationship between their magnitudes. It can be seen that RR has smaller variance, so RR suffices for binary graphs. The Laplace mechanism is mainly used for weighted graphs, since RR cannot handle weighted graphs. Our algorithms will provide both RR and Laplace versions, with RR for binary graphs and Laplace for weighted graphs.

\subsection{Triangle's One-Round Algorithm}
\label{sec:4.2}

According to \autoref{pro:3}, it can be deduced that the elements on the diagonal of the cube of the adjacency matrix \(A\) represent the number of ways a node can return to itself in exactly three steps. Since we assume there are no self-loops in the graph, any path that returns to the starting node in three steps must form a triangle. For each triangle, each node traverses the triangle in two directions, hence the total number of triangles \( f^{\bigtriangleup}\left( G \right)=\text{tr}\left( A^3 \right)/ (2 \cdot 3) \).

\begin{algorithm}[H]
\caption{TriOR}
\label{alg:2}
\begin{algorithmic}[1]
\Statex \textbf{Input:} Privacy budget \(\epsilon \in \mathbb{R}_{\geq 0} \), graph \(G \in \mathcal{G}\).
\Statex \textbf{Output:} Estimator: \(\hat{f}^{\bigtriangleup}\left( G \right)\).
\State \(\hat{A} \gets \text{GNAM}(G,\epsilon)\)
\State \(\hat{f}^{\bigtriangleup}\left( G \right) \gets \text{tr}\left( \hat{A}^3 \right) /6\)
\State \textbf{return} \( \hat{f}^{\bigtriangleup}\left( G \right) \)
\end{algorithmic}
\end{algorithm}

By the second property of \autoref{thm:1}, we can conclude that \(\mathbb{E} \left[ \text{tr}\left(\hat{A}^3 \right) \right] = \text{tr}\left(A^3 \right) \). Consequently, by directly computing \( \hat{f}^{\bigtriangleup}\left( G \right)=\text{tr}\left( \hat{A}^3 \right) /6 \), we can obtain an unbiased estimator of the number of triangles in the graph. Because there is only one interaction between users and data collector in GNAM, we denote this algorithm as Triangle's One-Round Algorithm (TriOR). \autoref{alg:2} presents the detailed implementation of TriOR. We can provide a closed-form MSE for TriOR:.

\begin{theorem}
\label{TriOR}
TriOR provides an unbiased estimator of \(f^{\bigtriangleup}\left( G \right)\), and its MSE equals to:
\begin{equation}
\sigma^2 \sum_{i < j} b_{ij}^2 + \sigma^4 (n-2)|E| + \frac{1}{6} \sigma^6 n(n-1)(n-2).
\end{equation}\label{eq:8}
\end{theorem}

In \autoref{app:Equ}, we prove the mathematical equivalence of TriOR, TCRR, and $\text{RR}_{\bigtriangleup}$. Consequently, the closed-form MSE expression we provide applies to all three algorithms, effectively refining the previously established MSE upper bounds. Furthermore, Eden et al. \cite{eden2025triangle} demonstrated that the MSE of TCRR matches the theoretical lower bound for non-interactive triangle counting algorithm, thereby proving the asymptotic optimality of the RR-based approach—a property that, by extension, is also shared by TriOR.

\subsection{Triangle's Two-Round Algorithm}

If a user \( v_u \) knows the relationships between their neighbors, they can count the number of triangles formed by themselves and their neighbors (denoted as \( f^{\bigtriangleup}_u \), which satisfies \(f^{\bigtriangleup}\left( G \right) = \frac{1}{3} \sum_{u=1}^{n} f^{\bigtriangleup}_u \)). The edges among neighbors is private information, which is not accessible to the user. However, users can estimate edge between neighbors based on the NAM, thus inferring the missing third edge in triangles and subsequently estimating \(f^{\bigtriangleup}\left( G \right)\).

\begin{algorithm}[H]
\caption{TriTR}
\label{alg:3}
\begin{algorithmic}[1]
\Statex \textbf{Input:} Privacy budget \(\epsilon \in \mathbb{R}_{\geq 0} \), graph \(G \in \mathcal{G}\).
\Statex \textbf{Output:} Estimator: \(\hat{f}^{\bigtriangleup}\left( G \right)\).
\Statex \#First round:
\State \(\hat{A} \gets \text{GNAM}(G,\epsilon)\)
\Statex \#Second round:
\For{\textbf{each} \textit{User} \(u=1\) to \(n\)}\textbf{:}
    \State \textbf{download} \(\hat{A}\)
    \State \(t_u \gets \sum_{\left( i,j \right):a_{ui}=a_{ju}=1}{\hat{a}_{ij}}\)
    \State \textbf{upload} \(t_u\) to \textit{Data Collector}
\EndFor
\Statex \textit{Data Collector} \textbf{do:}
\State \(\hat{f}^{\bigtriangleup}\left( G \right) \gets \frac{1}{6}\sum_{u=1}^n{t_u}\)
\State \textbf{return} \(\hat{f}^{\bigtriangleup}\left( G \right)\)
\end{algorithmic}
\end{algorithm}

\autoref{alg:3} shows the Triangle's Two-Round Algorithm (TriTR). In the first round, the data collector obtains the NAM through GNAM. In the second round, each user downloads the whole NAM \(\hat{A}\), computes \( t_u \) by summing the values between their neighbors in \(\hat{A}\), then uploads \( t_u \) to data collector. \( t_u \) satisfies \(\mathbb{E}[t_u] = 2 f^{\bigtriangleup}_u\), so \(\mathbb{E}[\frac{1}{6}\sum_{u=1}^n{t_u}]=f^{\bigtriangleup}\left( G \right)\). We now state its accuracy guarantee:

\begin{theorem}
\label{thm:4}
TriTR provides an unbiased estimator of \(f^{\bigtriangleup}\left( G \right)\), and its MSE equals to:
\begin{equation}
\frac{1}{9}\sigma^2 \sum_{i < j} b_{ij}^2.
\end{equation}
\end{theorem}

\subsection{Triangle's Modified Two-Round Algorithm}

TriTR has achieved a significant improvement in accuracy compared to TriOR. However, it faces the challenge of substantial download costs in the second round, where each user is required to download the entire NAM \(\hat{A}\). To address this issue, we adopt a trade-off approach, which sacrifices slightly in accuracy but significantly reduces the download cost from the entire matrix to just one column.

\begin{algorithm}[H]
\caption{TriMTR}
\label{alg:4}
\begin{algorithmic}[1]
\Statex \textbf{Input:} Privacy budget \(\epsilon \in \mathbb{R}_{\geq 0} \), graph \(G \in \mathcal{G}\).
\Statex \textbf{Output:} Estimator: \(\hat{f}^{\bigtriangleup}\left( G \right)\).
\Statex \#First round:
\State \(\hat{A} \gets \text{GNAM}(G,\epsilon)\)
\State \textit{Data Collector} \textbf{compute:} \(\hat{B} \gets \hat{A}^2\)
\Statex \#Second round:
\For{\textbf{each} \textit{User} \(u=1\) to \(n\)}\textbf{:}
    \State \textbf{download} the \(u\)-th column of \(\hat{B}\)
    \State \(t_u \gets \sum_{i: a_{ui}=1}{\hat{b}_{iu}}\)
    \State \textbf{upload} \(t_u\) to \textit{Data Collector}
\EndFor
\Statex \textit{Data Collector} \textbf{do:}
\State \(\hat{f}^{\bigtriangleup}\left( G \right) \gets \frac{1}{6}\sum_{u=1}^n{t_u}\)
\State \textbf{return} \(\hat{f}^{\bigtriangleup}\left( G \right)\)
\end{algorithmic}
\end{algorithm}

\autoref{alg:4} shows the Triangle's Modified Two-Round Algorithm (TriMTR). In the first round, the data collector computes \(\hat{B} = \hat{A}^2\) after GNAM. In the second round, each user \(v_u\) downloads the \(u\)-th column of \(\hat{B}\), sums the neighbor-associated entries in \(t_u\). According to the first property of \autoref{thm:1}, \(\mathbb{E}[\hat{b}_{iu}] = b_{iu}\). Therefore, if \(a_{ui} = 1\), then \(\mathbb{E}[a_{ui}\hat{b}_{iu}] = a_{ui}b_{iu}=b_{iu}\), which represents the number of triangles formed by the edge \((v_u, v_i)\). And thus \(\mathbb{E}[t_u] = 2f^{\bigtriangleup}_u\). Consequently, \(\mathbb{E}\left[\frac{1}{6}\sum_{u=1}^n t_u\right] = f^{\bigtriangleup}(G)\). Then we show its accuracy guarantee:

\begin{theorem}
\label{thm:5}
TriMTR provides an unbiased estimator of \(f^{\bigtriangleup}\left( G \right)\), and its MSE equals to:
\begin{equation}
\frac{4}{9} \sigma^2 \sum_{i < j} b_{ij}^2 + \frac{1}{9} \sigma^4 (n-2)|E|.
\end{equation}
\end{theorem}

\subsection{Quadrangle's Two-Round Algorithm}

TriTR uses \(\hat{A}\) to complete the missing one edge. Similarly, we can employ \(\hat{B}\) in TriMTR for missing two edges. Let \(f^{\Box}_u\) denote the number of quadrangles formed by \(v_u\), which satisfies \(f^{\Box}(G) = \frac{1}{4}\sum_{u=1}^n f^{\Box}_u\). Following this intuition, we designed Quadrangle's Two-Round Algorithm (QuaTR).

\begin{algorithm}[H]
\caption{QuaTR}
\label{alg:5}
\begin{algorithmic}[1]
\Statex \textbf{Input:} Privacy budget \(\epsilon \in \mathbb{R}_{\geq 0} \), graph \(G \in \mathcal{G}\).
\Statex \textbf{Output:} Estimator: \(\hat{f}^{\Box}\left( G \right)\).
\Statex \#First round:
\State \(\hat{A} \gets \text{GNAM}(G,\epsilon)\)
\State \textit{Data Collector} \textbf{compute:} \(\hat{B} \gets \hat{A}^2\)
\Statex \#Second round:
\For{\textbf{each} \textit{User} \(u=1\) to \(n\)}\textbf{:}
    \State \textbf{download} \(\hat{B}\)
    \State \(q_u \gets \sum_{\left( i,j \right):a_{ui}=a_{ju}=1}{\left( \hat{b}_{ij}-1 \right)}\)
    \State \textbf{upload} \(q_u\) to \textit{Data Collector}
\EndFor
\Statex \textit{Data Collector} \textbf{do:}
\State \(\hat{f}^{\Box}\left( G \right) \gets \frac{1}{8}\sum_{u=1}^n{q_u}\)
\State \textbf{return} \(\hat{f}^{\Box}\left( G \right)\)
\end{algorithmic}
\end{algorithm}

\autoref{alg:5} shows the implementation of QuaTR, which is similar to that of TriTR, with modifications only in lines 5 and 7. The change in line 5 is necessary because user \(v_u\) provides two edges \(v_j \to v_u\) and \(v_u \to v_i\), while the remaining paths of the form \(v_i \to v_k \to v_j\) are provided by \(\hat{b}_{ij}\). Since \(\mathbb{E}[\hat{b}_{ij}]=b_{ij}\), and \(b_{ij}\) includes the path \(v_i \to v_u \to v_j\), which cannot form a quadrangle with \(v_j \to v_u \to v_i\), it is necessary to subtract 1 to exclude this invalid path. Since each of the four vertices of a quadrangle will traverse the quadrangle from two distinct directions, we need to change line 7 by dividing by \(2 \times 4=8\). Then we show its accuracy guarantee:

\begin{theorem}
\label{thm:6}
QuaTR provides an unbiased estimator of \(f^{\Box}\left( G \right)\), and its MSE equals to:
\begin{equation}
\frac{1}{4}\sigma^2 \sum_{i<j} c_{ij}^2 + \frac{1}{16}(n-2)\sigma^4 \sum_{i<j} b_{ij}^2
\end{equation}
\end{theorem}

\subsection{Summary}

\begin{figure}[h]
  \centering
  \includegraphics[width=\linewidth]{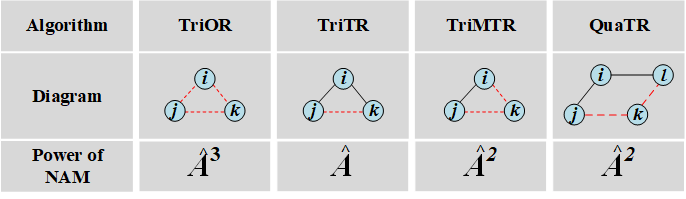}
  \caption{\textbf{An Intuitive Overview of the Algebraic Structure of Our Proposed Algorithms.}}
  \label{fig:3}
\end{figure}

The intuitive ideas behind the design of these algorithms are illustrated in \autoref{fig:3}. Black solid edges denote true edges in the original graph, whereas red dashed edges denote noisy edges estimate from NAM \(\hat{A}\). The powers of the noisy adjacency matrix correspond to the number of noisy edges within paths of a given length. 

Moreover, the variance of each algorithm can be expressed using powers of the  adjacency matrix, revealing a close connection between our estimator designs and matrix structure. 
For a matrix $A\in\mathbb{R}^{m\times n}$, the Frobenius norm is
\[
\|A\|_F \triangleq \sqrt{\sum_{i=1}^{m}\sum_{j=1}^{n} a_{ij}^2 } .
\]
Define the $0$-th power of any matrix $M\in\mathbb{R}^{n\times n}$ as $M^0 \triangleq I_n$. 
Ignoring constant factors, we obtain the following patterns:

\begin{equation}
\mathbb{V}(\text{TriTR})=O(\sigma^2 \|A^2\|^2_F)
\label{eq:12}
\end{equation}
\begin{equation}
\mathbb{V}(\text{TriMTR})=O(\sigma^2 \|A^2\|^2_F + n \sigma^4 \|A^1\|^2_F)
\end{equation}
\begin{equation}
\mathbb{V}(\text{TriOR})=O(\sigma^2 \|A^2\|^2_F + n \sigma^4 \|A^1\|^2_F + n^2 \sigma^6 \|A^0\|^2_F)
\end{equation}
\begin{equation}
\mathbb{V}(\text{QuaTR})=O(\sigma^2 \|A^3\|^2_F + n \sigma^4 \|A^2\|^2_F)
\label{eq:15}
\end{equation}

We observe that for a \(k\)-loop algorithm, the variance includes at least a term of order 
\( O(\sigma^2 \|A^{k-1}\|_F^2) \). Moreover, each additional noisy edge contributes an extra variance term: relative to the previous term, it introduces an additional factor of \(n\sigma^2\) while decreasing the matrix power inside the Frobenius norm \(\|\cdot\|^2_F\) by one.

\textbf{Directed Graph Application.} In directed graphs where edges have directionality, the adjacency matrix becomes asymmetric. Since \autoref{thm:1} remains valid for directed graphs, only minor modifications are needed to adapt these algorithms to the directed case. For directed graphs, defining triangles and quadrangles as 3-step and 4-step loops, respectively. Modification of GNAM involves removing line 3 and changing line 6 to: \(\tilde{A} \gets L\). For TriOR, TriTR and TriMTR, adjust the final step by changing the coefficient: \(\hat{f}^{\bigtriangleup}\left( G \right) \gets \frac{1}{3}\sum_{u=1}^n{t_u}\) (modified from \(\frac{1}{6}\) to \(\frac{1}{3}\)). For QuaTR, the modifications involve keeping \(\hat{b}_{ij}\) unchanged in line 5 (without subtracting 1) and adjusting the final coefficient from \(\frac{1}{8}\) to \(\frac{1}{4}\). The modified algorithms maintain the same variance form as shown in \autoref{eq:12}--\ref{eq:15}.

\section{Differential Privacy on Randomized Data}
\label{sec:5}

Constructing high-accuracy interactive protocols requires resolving a fundamental theoretical challenge: preserving differential privacy when the second-round statistic depends on randomized data produced in previous rounds.

To address the above issues, we propose the \textbf{Differential Privacy on Randomized Data} (DPRD) framework and prove that it can be converted to standard DP via a composition theorem, and that its definition even subsumes that of DP. We then introduce two fundamental tools in graph DP: Graph Projection and Clipping. Finally, we instantiate DPRD to revise the two-round algorithms in \autoref{sec:5.4}-\ref{sec:5.5}, obtaining both pure \(\epsilon\)-LDP versions and standard \((\epsilon,\delta)\)-LDP version.

\subsection{Key Challenge of Interactive DP}
\label{sec:5.1}

In a multi-round LDP protocol, the input of a later round may depend on randomized outputs from previous rounds. This creates a non-trivial privacy challenge for the second round. Standard DP mechanisms, such as the Laplace mechanism, usually require a deterministic sensitivity \(\Delta f\). However, when the second-round statistic depends on a randomized object, its sensitivity may also become random or difficult to compute.

A naive approach of directly releasing the second-round statistic poses a severe privacy risk. For example, when GNAM uses continuous noise, the noisy matrix \(\hat{A}\) almost surely takes different values in different executions. The unperturbed second-round statistic may then become a fingerprint of the user's neighborhood. An attacker could enumerate possible neighbor lists and match the resulting second-round values with the observed value, thereby inferring the user's private adjacency list.

Therefore, the second round must provide a rigorous privacy guarantee. The key question is how to correctly calibrate the second-round noise when its input is computed on randomized data.

\subsection{Definition and Property of DPRD}
\label{sec:5.2}

We first formulate the problem and provide the definition of \((\epsilon,\delta)\)-Differential Privacy on Randomized Data in \autoref{def:DPRD}. The key point is that this definition depends on an already observed randomized data \(\xi\). In the setting of this paper, \(D\) can be understood as the adjacency list \(\mathbf{a}_u\) of user \(u\), while \(\xi\) corresponds to the NAM obtained by GNAM in the first round.

\begin{definition}[\((\epsilon,\delta)\)-DPRD]
\label{def:DPRD}
Let \(f:\mathcal{X}^n\times\Xi\to\mathcal{Y}\) be a measurable operator, let \(P\) be a distribution over \(\Xi\), and let \(\xi\sim P\) be randomized data. Let \(\mathcal{R}:\mathcal{Y}\to\mathcal{Z}\) be a randomized mechanism. For each dataset \(D\in\mathcal{X}^n\), define \(\mathcal{M}_{P}(D)=\mathcal{R}(f(D,\xi))\). We say that \(\mathcal{M}_{P}\) satisfies \((\epsilon,\delta)\)-differential privacy on randomized data with respect to \(P\), if for any neighboring datasets \(D,D'\in\mathcal{X}^n\) and any measurable family of output events \(\{S_{\eta}\subseteq\mathcal{Z}:\eta\in\Xi\}\),
\[
\Pr_{\xi\sim P,\mathcal{R}}
\left[
\mathcal{R}(f(D,\xi))\in S_{\xi}
\right]
\leq
e^{\epsilon}
\Pr_{\xi\sim P,\mathcal{R}}
\left[
\mathcal{R}(f(D',\xi))\in S_{\xi}
\right]
+\delta.
\]
\end{definition}

Just providing the definition of DPRD is not sufficient, it is meaningful only if it can be embedded into the standard DP framework, so that its practical use can be examined under existing DP principles. To this end, we establish the composition theorem between DPRD and DP, which shows that DPRD can be converted into standard DP.

\begin{theorem}
\label{thm:dprd}
Let \(\mathcal{R}_1: \mathcal{X}^n \to \Xi\) be an \((\varepsilon_1,\delta_1)\)-differentially private mechanism that maps a dataset \(D\in\mathcal{X}^n\) to a randomized data point \(\xi_D=\mathcal{R}_1(D)\). Let \(P_D\) denote the distribution of \(\xi_D\) (because it depends on \(D\)). 
Let \(f:\mathcal{X}^n\times\Xi\to\mathcal{Y}\) be a measurable operator, and let \(\mathcal{R}_2 \circ f :\mathcal{X}^n\times\Xi \to\mathcal{Z}\) be a mechanism that satisfies \((\varepsilon_2,\delta_2)\)-differential privacy on randomized data with respect to \(P_D\). 
Then the composed mechanism
\[
\mathcal{M}(D)=\bigl(\xi_D,\;\mathcal{R}_2(f(D,\xi_D))\bigr)
\]
preserves \((\varepsilon_1+\varepsilon_2,\;\delta_1+\delta_2)\)-differential privacy (\autoref{def:1}).
\end{theorem}

If the distribution of \(\xi\) is independent of \(D\), then \(\xi\) can be viewed as the output of a \((0,0)\)-DP mechanism. Therefore, this setting can be regarded as a special case of \autoref{thm:dprd}, which implies that standard \((\epsilon,\delta)\)-DP is recovered as a degenerate instance of DPRD when the auxiliary randomness is independent of the data.

\subsection{Graph Projection and Clipping}
\label{sec:5.3}

Graph Projection~\cite{day2016publishing, kasiviswanathan2013analyzing, raskhodnikova2015efficient, imola2022differentially} and Clipping~\cite{dwork2006calibrating, dwork2014algorithmic} are two basic tools in graph DP and DP, respectively. Since both tools will be used in our subsequent algorithms, we briefly introduce them in this subsection.

\textbf{Graph Projection.} The main purpose of graph projection is to obtain the noisy degree of each node and ensure, through \textit{edge removal}, that the degree after graph projection satisfies \(d_u \leq \tilde{d}_u\). This condition is essential because it ensures that, in subsequent operations, the sensitivity of the data uploaded by user \(u\) is controlled by \(\tilde{d}_u\). If the degree exceeds \(\tilde{d}_u\), the sensitivity may be underestimated, and DP may no longer be guaranteed. Since \textit{edge removal} deletes some edges from the original graph, it effectively maps the original graph to a graph with fewer edges. As a result, the number of subgraphs in the projected graph may decrease, making the subsequent estimator biased~\cite{imola2022communication, he2025robust}.

\begin{algorithm}[H]
\caption{Graph Projection}
\label{alg:6}
\begin{algorithmic}[1]
\Statex \textbf{Input:} Adjacency list \(\mathbf{a}_i\), degree \(d_i\), privacy budget \(\epsilon_0\), parameter \(\alpha\).
\Statex \textbf{Output:} Processed adjacency list \(\mathbf{a}'_i\), noisy degree \(\tilde{d}_i\).
\State \(\tilde{d}_i =\lfloor \alpha+\max \left\{ d_i + \text{Lap}\left(\frac{1}{\epsilon_0}\right), 0\right\}\rfloor \)
\State Remove \(d_i-\tilde{d}_i\) neighbors randomly if \(\tilde{d}_i < d_i\) to get \(\mathbf{a}'_i\).
\State \textbf{return} \(\mathbf{a}'_i\), \(\tilde{d}_i\)
\end{algorithmic}
\end{algorithm}

\autoref{alg:6} presents the graph projection procedure used in this work. We slightly modify the version in~\cite{imola2022differentially} by ensuring that \(\tilde{d}_u \geq \alpha\) for any \(u\). Therefore, under the same choice of \(\alpha\), our graph projection avoids removing edges incident to low-degree nodes, thereby reducing the probability of \textit{edge removal} and hence the resulting bias. For nodes with \(d_u < \alpha\), \textit{edge removal} never occurs. For nodes with \(d_u \geq \alpha\), the probability of \textit{edge removal} is \(\frac{1}{2}e^{-\epsilon_0 \alpha}\). For instance, when \(\alpha=150\) and \(\epsilon_0=0.1\), this probability evaluates to \(1.53\times 10^{-7}\), implying that the resulting bias is negligible. However, it should be noted that a larger \(\alpha\) is not always better, because the sensitivity of subsequent operations depends on \(\tilde{d}_u\). If \(\tilde{d}_u\) is too large, the subsequent sensitivity will also increase, leading to additional noise.

\textbf{Clipping.} Clipping is a widely used tool in DP. For any \(\kappa>0\), define
\(
\text{clip}(X,\kappa)
=\max\{\min\{X,\kappa\},-\kappa\}.
\)
This operation is referred to as clipping. It restricts all values to the range \([-\kappa,\kappa]\), which makes it convenient to combine with the Laplace mechanism for DP protection. Clipping also introduces bias. To address this issue, we prove \autoref{pro:clip}, which shows that relating \(\kappa\) to the \(\beta\)-quantile of the distribution can effectively control the bias by controlling the value of \(\beta\). The applicability of \autoref{pro:clip} encompasses both the Laplace distribution and the Gaussian distribution.

\begin{proposition}
\label{pro:clip}
Let $X$ be a continuous random variable with finite expectation $\mu = \mathbb{E}[X]$, and suppose the distribution of $X$ is symmetric about $\mu$. For any $\beta \in [0, 0.5]$, denote the upper and lower $\beta$-quantiles of $X$ as $q_{\beta}^+$ and $q_{\beta}^-$, respectively. Let $\kappa_{\beta} = \max\{\|q_{\beta}^+\|, \|q_{\beta}^-\|\}$. The relative bias introduced by clipping then satisfies:
\begin{equation}
0 \leq \frac{\mathbb{E}[X] - \mathbb{E}\left[\operatorname{clip}\left(X, \kappa_{\beta}\right)\right]}{\mathbb{E}[X]} \leq 2\beta.
\end{equation}
\end{proposition}

\subsection{Revisions to Two-Round Algorithms}
\label{sec:5.4}

\textbf{Key Idea.} We use clipping to control the sensitivity in the second round, and then apply the Laplace mechanism to preserve \(\epsilon_2\)-DPRD on NAM \(\hat{A}\) and noisy degree \(\tilde{d}_u\). Then, by \autoref{thm:dprd}, we can prove that the whole process preserves (\(\epsilon_0+\epsilon_1+\epsilon_2\))-edge LDP. The bias introduced by clipping is controlled by setting the parameter \(\beta\) in \autoref{pro:clip}.

\begin{algorithm}[H]
\caption{Second Randomizer of \(\text{TriTR}\)}
\label{alg:7}
\begin{algorithmic}[1]
\Statex \textbf{Input:} Noisy adjacency matrix \(\hat{A}\), user \(v_u\)'s noisy degree \(\tilde{d}_u\), \(v_u\)'s neighbors \(\mathbf{adj}_u\), privacy budget \(\epsilon_2\), threshold \(\beta\).
\Statex \textbf{Output:} \(t_u\).
\State \(\kappa \gets \frac{e^{\epsilon_1}}{e^{\epsilon_1}-1}\text{ (RR)} \text{ or }1+\frac{1}{\epsilon_1}\ln\frac{1}{2\beta} \text{ (Laplace Mechanism)}\)
\State \(sum_u \gets \sum_{\left( i,j \right):a_{ui}=a_{ju}=1}{\text{clip}(\hat{a}_{ij},\kappa)}\)
\State \(t_u \gets sum_u + \text{Lap}\left(\frac{2\kappa\tilde{d}_u}{\epsilon_2} \right)\)
\State \textbf{return} \(t_u\)
\end{algorithmic}
\end{algorithm}

\begin{figure}[h]
\centering
\includegraphics[width=0.8\linewidth]{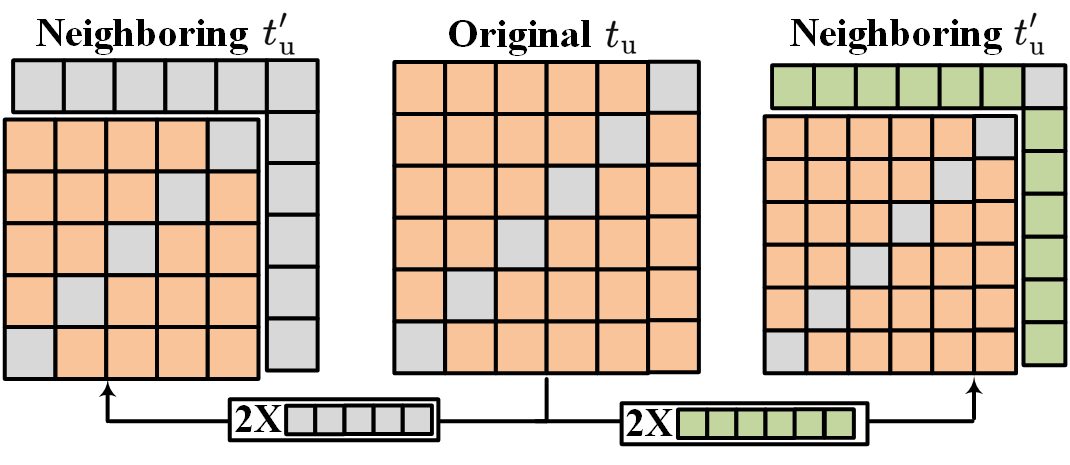}
\caption{\textbf{Illustrating the Change Distribution in DPRD.} When a user changes a single edge, e.g., adds or removes a neighbor \(z\), the local statistic \(t_u\) becomes a neighboring statistic \(t_u'\). The change \(\max\|t_u' - t_u\|_1\) is the sensitivity.}
\label{fig:4}
\end{figure}

\textbf{Revision to TriTR (\autoref{alg:7}).} Due to the randomness of \(\hat{A}\), the \(t_u\) and \(t_u'\) are essentially sums of a series of random variables, which can be viewed as random variables following different distributions. \autoref{fig:4} visually shows the difference between \(t_u\) and \(t_u'\). 

Since each element in \(\hat{A}\) either follows a binary distribution or a Laplace distribution, we can set an upper bound \(\kappa\) on the absolute value of each element to limit the change in sensitivity \(\max\|t_u' - t_u\|_1\). The final sensitivity has a deterministic upper bound \(2\kappa \tilde{d}_u\). Therefore, adding noise from \(\text{Lap}\left(\frac{2\kappa\tilde{d}_u}{\epsilon_2} \right)\) suffices to preserve \(\epsilon_2\)-DPRD on \((\hat{A}, \tilde{d}_u)\) in the second round.

If RR is chosen as the first randomizer in GNAM, since the elements in \(\hat{A}\) follow a binary distribution, we directly set \(\kappa\) to the larger absolute value of the two points. If Laplace is chosen in GNAM, we set \(\kappa\) to the upper \(\beta\)-quantile of \(\text{Lap}(1, 1/\epsilon_1)\) and then apply clipping. The bias introduced by clipping can be controlled by \autoref{pro:clip}.

\begin{algorithm}[H]
\caption{Second Randomizer of TriMTR}
\label{alg:8}
\begin{algorithmic}[1]
\Statex \textbf{Input:} Noisy two-step matrix \(\hat{B}\), user \(v_u\)'s noisy degree \(\tilde{d}_u\), \(v_u\)'s neighbors \(\mathbf{adj}_u\), privacy budget \(\epsilon_2\), threshold \(\beta\).
\Statex \textbf{Output:} \(t_u\).
\State \(\kappa_u \gets \Phi^{-1}(1 - \beta) \cdot \sqrt{(n-2)\sigma^4+\left(\tilde{d}_u+\tilde{d}_\max\right)\sigma^2} + \tilde{d}_u \)
\State \(sum_u\gets \sum_{i\in \mathbf{adj}_u}{\text{clip}(\hat{b}_{iu}, \kappa_u)}\)
\State \(t_u \gets sum_u + \text{Lap}(\kappa_u/\epsilon_2) \)
\State \textbf{return} \(t_u\)
\end{algorithmic}
\end{algorithm}

\textbf{Revision to TriMTR (\autoref{alg:8}).} Similar to TriTR, \(t_u\) and \(t_u'\) are still random variables from two different distributions. However, the sensitivity \(\max\|t_u' - t_u\|_1\) becomes much simpler: it is the maximum absolute value among the individual elements in the \(u\)-th column of \(\hat{B}\), i.e., \(\hat{b}_{iu}\). 

Due to the randomness of \(\hat{B}\), sensitivity \(\max\|t_u' - t_u\|_1\) may take extremely large values, which would increase the noise added in the second round. Therefore, we still apply clipping to sacrifice some unbiasedness in order to significantly reduce the variance. Since \(\hat{b}_{ij} = \sum_{k=1}^n a_{ik} a_{kj}\), which is a sum of \(n\) independent terms \(a_{ik} a_{kj}\), by the \textit{Central Limit Theorem}, \(\hat{b}_{ij}\) is approximately distributed as \(\mathcal{N}\left(b_{ij}, (n-2)\sigma^4 + (d_i + d_j)\sigma^2 \right)\). Hence, we set \(\kappa_u\) as possible maximum value of upper \(\beta\)-quantile of \(\hat{b}_{iu}\).

\begin{algorithm}[H]
\caption{Second Randomizer of \(\text{QuaTR}\)}
\label{alg:9}
\begin{algorithmic}[1]
\Statex \textbf{Input:} Noisy two-step matrix \(\hat{B}\), user \(v_u\)'s noisy degree \(\tilde{d}_u\), \(v_u\)'s neighbors \(\mathbf{adj}_u\), privacy budget \(\epsilon_2\), threshold \(\beta\).
\Statex \textbf{Output:} \(q_u\).
\State \(\kappa \gets \Phi^{-1}(1 - \beta) \cdot \sqrt{(n-2)\sigma^4+2\tilde{d}_\max \sigma^2} + \tilde{d}_\max -1 \)
\State \(sum_u \gets \sum_{\left( i,j \right):a_{ui}=a_{ju}=1}{ \text{clip}(\hat{b}_{ij}-1,\kappa) }\)
\State \(q_u \gets sum_u + \text{Lap}\left(\frac{2\kappa\tilde{d}_u}{\epsilon_2} \right)\)
\State \textbf{return} \(q_u\)
\end{algorithmic}
\end{algorithm}

\textbf{Revision to QuaTR (\autoref{alg:9}).} Similar to TriTR, \(q_u\) and \(q_u'\) are still random variables from two different distributions, and even the form of the sensitivity \(\max\|q_u' - q_u\|_1\) resembles that of TriTR. The main difference is that the sensitivity in TriTR is a sum of several \(\hat{a}_{ij}\) terms, while in QuaTR it is a sum of several \((\hat{b}_{ij} - 1)\) terms. 

Since \(\hat{b}_{ij}\) approximately follows a Gaussian distribution as shown in TriMTR, we can adopt the same approach to set an upper bound \(\kappa\) on the absolute value of each \((\hat{b}_{ij}-1)\). Then, similar to TriTR, the total sensitivity is given as \(2\kappa \tilde{d}_u\).

\textbf{DP guarantee.} In the specific algorithm procedure, in the first round, each node first processes its own adjacency list using graph projection, then executes its part of GNAM, and finally uploads both \(\tilde{\mathbf{a}}_u\) and \(\tilde{d}_u\) to the data collector simultaneously. In the second round, we replace the second-round parts in \autoref{alg:3}--\ref{alg:5} with \autoref{alg:7}--\ref{alg:9}, specifically: line 4--5 of TriTR, line 5--6 of TriMTR, and line 5--6 of QuaTR. After the aforementioned modifications to the two-round algorithms, the following theorem holds for all of them.

\begin{theorem}
\label{thm:2RPureDP}
Substituting \autoref{alg:7}--\ref{alg:9} into the corresponding positions of \autoref{alg:3}--\ref{alg:5}, the resulting algorithms preserve \((\epsilon_0+\epsilon_1+\epsilon_2)\)-edge LDP and \(2(\epsilon_0+\epsilon_1+\epsilon_2)\)-edge DP.
\end{theorem}

\subsection{Variance Reduction for TriTR}
\label{sec:5.5}

Since \autoref{alg:7} clips each random variable with an upper bound \(\kappa\), and because \(\|t_u' - t_u\|_1\) contains a total of \(2\tilde{d}_u\) elements, the resulting sensitivity is \(2\kappa \tilde{d}_u\). Statistically speaking, the sum of the upper \(\beta\)-quantiles of several random variables is much larger than the upper \(\beta\)-quantile of the sum of those random variables. Therefore, the sensitivity obtained in \autoref{alg:7} is overly large, which leads to excessive noise being added in the second round.

Thus, a natural idea emerges: we directly compute the upper \(\beta\)-quantile of \(\|t_u' - t_u\|_1\), set it as \(\kappa\), and then clip the row sums and column sums of \(t_u\) to obtain a tighter sensitivity bound, after which we add Laplace noise to preserve \((\epsilon_0 + \epsilon_1 + \epsilon_2)\)-edge LDP. \textbf{However}, our study shows that one can always construct a specific \(\hat{A}\) such that, although each row sum and each column sum does not exceed \(\kappa\), the total \(\|t_u' - t_u\|_1\) still exceeds \(2\kappa\). Due to the randomness of \(\hat{A}\), this implies that such an approach cannot preserve \((\epsilon_0 + \epsilon_1 + \epsilon_2)\)-edge LDP.

\textbf{Key Idea.} Since it is impossible to guarantee that the second round preserves \(\epsilon_2\)-DPRD on \((\hat{A}, \tilde{d}_u)\), we can design an algorithm that preserves \((\epsilon_2, \delta)\)-DPRD on \((\hat{A}, \tilde{d}_u)\) instead. As long as \(\delta\) is small, typically \(\delta \ll 1/n\), we can still provide strong privacy protection. The key point is that we abandon clipping and allow some row sums or column sums to exceed the sensitivity bound, but the probability of such exceedance is controlled by \(\delta\).

To distinguish it from the approach implemented in \autoref{alg:7}, we denote the new algorithm obtained from the complete procedure as \(\text{TriTR}^*\).

\begin{algorithm}[H]
\caption{Second Randomizer of \(\text{TriTR}^*\)}
\label{alg:10}
\begin{algorithmic}[1]
\Statex \textbf{Input:} Noisy adjacency matrix \(\hat{A}\), user \(v_u\)'s noisy degree \(\tilde{d}_u\), privacy budget \(\epsilon_1, \epsilon_2, \delta\).
\Statex \textbf{Output:} \(t_u\).
\State \(v \gets \frac{e^{\epsilon_1}}{3(e^{\epsilon_1}-1)}\text{ (RR)} \text{ or } \frac{1}{\epsilon_1} \text{ (Laplace Mechanism)}\)
\State \(\kappa_u \gets \tilde{d}_u + v \ln\frac{2 \tilde{d}_u }{\delta} + \sqrt{ v^2 \ln^2\frac{2 \tilde{d}_u}{ \delta} + 2\tilde{d}_u \sigma^2\ln\frac{2 \tilde{d}_u }{\delta} } \)
\State \(t_u \gets \sum_{\left( i,j \right):a_{ui}=a_{ju}=1}{\hat{a}_{ij}} + \text{Lap}\left(\frac{2\kappa_u}{\epsilon_2} \right)\)
\State \textbf{return} \(t_u\)
\end{algorithmic}
\end{algorithm}

\textbf{Further Revision to TriTR (\autoref{alg:10}).} If the probability that each absolute value of row sum of \(t_u\) exceeds \(\kappa_u\) is less than \(\frac{1}{d_u}\delta\), then by the symmetry of \(\hat{A}\), each column sum also satisfies the same property. Define the above event as the \textit{Good Event}. Then the probability that the \textit{Good Event} occurs is at least \(1 - \delta\). It can be shown that, conditioned on the \textit{Good Event}, if the algorithm preserves \(\epsilon_2\)-DP, then it preserves \((\epsilon_2, \delta)\)-DPRD on \((\hat{A}, \tilde{d}_u)\). By \autoref{thm:dprd}, we can prove that the entire procedure preserves \((\epsilon_0 + \epsilon_1 + \epsilon_2, \delta)\)-edge LDP.

To ensure that the probability that the absolute value of a row sum exceeds \(\kappa_u\) is less than \(\frac{1}{d_u}\delta\), it suffices to make the probabilities of exceeding the upper bound and the lower bound each less than \(\frac{1}{2d_u}\delta\). Then, we apply \textit{Bernstein's inequality} to obtain the explicit expression for \(\kappa_u\). Finally, since the true value of \(d_u\) is unknown, we take the possible maximum \(\tilde{d}_u\). More detailed derivations and the proof are given in the proof of \autoref{thm:epsilon_delta_DP} (\autoref{sec:E}).

\begin{theorem}
\label{thm:epsilon_delta_DP}
After substituting \autoref{alg:10} with the line 4--5 of \autoref{alg:3}, the resulting algorithms preserve standard \((\epsilon_0+\epsilon_1+\epsilon_2,\delta)\)-edge LDP and \((2(\epsilon_0+\epsilon_1+\epsilon_2),2\delta)\)-edge DP.
\end{theorem}

\subsection{Utility Discussion}
\label{sec:5.6}

\textbf{MSE Guarantee.} For two-round algorithms, the MSE consists of bias and variance, but the bias is nearly negligible relative to variance. The bias sources are edge removal from graph projection and truncation from clipping. The edge removal probability is strictly controlled by \(\alpha\), and the clipping bias is controlled by \(\beta\) as in \autoref{pro:clip}; both can be easily made very small (also confirmed in experiments). Hence, the MSE is dominated by variance, which can be split into noise from the first and second rounds. Since these two noise parts are independent, total variance is their sum. We then prove an upper bound on MSE in \autoref{pro:MSE}, and compare our algorithm with existing ones in \autoref{tab:1}.

\begin{proposition}
\label{pro:MSE}
Let \(\text{MSE}_1\), \(\text{MSE}_2\), \(\text{MSE}_3\) and \(\text{MSE}_4\) denote the mean squared errors of TriTR, TriMTR, QuaTR and \(\text{TriTR}^*\), respectively. After substituting \autoref{alg:7}--\ref{alg:10} into the corresponding positions of \autoref{alg:3}--\ref{alg:5}, the following asymptotic upper bounds hold:
\begin{equation}
\text{MSE}_1\leqslant O\left(nd_{\max}^3\right), \quad \text{MSE}_2\leqslant O\left(nd_{\max}^3 + n^2d_{\max}\right),
\end{equation}
\begin{equation}
\text{MSE}_3 \leqslant O\left(nd_{\max}^5 + n^2d_{\max}^3\right), \quad \text{MSE}_4\leqslant O\left(nd_{\max}^3\right).
\end{equation}
\end{proposition}

\textbf{RE Guarantee.} MSE is convenient for mathematical analysis, but in practice, RE is more relevant. We also provide theoretical guarantees on RE for TriTR, TriMTR, and \(\text{TriTR}^*\). To the best of our knowledge, we are the first to give theoretical guarantees on RE for private subgraph counting. 

To do so, we introduce a commonly used graph statistic: the clustering coefficient \(C_g = \frac{3 \times \# \text{triangles}}{\#\text{2-stars}}\), which is one of the typical quantities estimated in private subgraph counting. Using the clustering coefficient, we derive upper bounds on the RE for some of the second-round algorithms, as stated in \autoref{thm:TriRE}. As long as the graph contains at least one triangle, \autoref{equ:RE} is well-defined.

\begin{theorem}
\label{thm:TriRE}
If \(C_g>0\), the RE of TriTR, TriMTR and \(\text{TriTR}^*\) satisfy
\begin{equation}
\mathrm{RE} \le O\left( \frac{1}{C_g d_{\mathrm{avg}}} \right).
\label{equ:RE}
\end{equation}
\end{theorem}

\textbf{Discussion.} \autoref{thm:TriRE} reveals a key property: if the clustering coefficient \(C_g\) and average degree \(d_{avg}\) remain stable, the relative error of our two-round algorithms does not grow with network size \(n\). The RE mainly depends on \(C_g\) and \(d_{avg}\); denser graphs give higher accuracy. In practice, as \(n\) increases, \(d_{avg}\) often also increases, so accuracy may improve or stay stable at larger scales. In contrast, the one-round algorithm TriOR lacks this property because its variance contains an excessively large \(O(n^3)\) term. This further demonstrates the advantage of two-round algorithms over one-round ones for large-scale decentralized graph analytics.

\textbf{TriMTR is Our Favorite Algorithm.} Its RE remains essentially the same as TriTR and has the same relative-err upper bound, but lowers per-node download cost to $O(n)$. The data collector handles extra computation (\(O(n^3)\)), which is acceptable versus TriTR's huge communication cost (\(O(n^2)\)). For $n=10^6$, with a 5090-class GPU, the collector's computation finishes within one day (\autoref{sec:D}), usually less than the time to gather data from $n$ users. TriTR, however, requires 58.2 GiB per user (vs. 3.8 MiB for TriMTR), which is impractical. Moreover, TriMTR offloads computation to the collector. Per-user cost is $O(n)$ for TriMTR, but $O(n + d_u^2)$ for TriTR, which hurts users with large $d_u$.

\section{Experimental Evaluation}

In this section, we validate the proposed algorithms using real-world datasets to address the following three questions:

\textbf{RQ1.} Do the experimental results align with theoretical derivations, and how do our algorithms compare against state-of-the-art baselines under different parameter settings?

\textbf{RQ2.} In the two-round algorithms, two parameters \(\alpha\) and \(\beta\) are introduced for graph projection and clipping. How should these be set in practice?

\textbf{RQ3.} Why is the bias introduced by graph projection and clipping in experiments almost negligible?

Experimental code and results are open-sourced at~\cite{anonymous2026SubGraph_DP_NAM}.

\subsection{Experimental Set-up}
\label{sec:6.1}

\textbf{Datasets.} We use two real graph datasets from~\cite{snapnets}: \textbf{Facebook}~\cite{leskovec2012learning} (4,039 nodes, 88,234 edges) and \textbf{CA-AstroPh}~\cite{leskovec2007graph} (18,772 nodes, 198,110 edges). Both of them can be downloaded at~\cite{snapnets}.

\textbf{Experimental Environment.} We evaluate each algorithm by its average relative error over 50 runs on an NVIDIA RTX A6000 GPU.

\textbf{For RQ1.} Following the convention~\cite{imola2022differentially}, we set \(\epsilon=1\) and \(\delta=1/(100n)\). For \(\text{WShuffle}_{\bigtriangleup}^*\), we set \(c=1\) and allocate \(\epsilon\) with ratio \(0.1:0.9\)~\cite{imola2022differentially}. We evaluate two-round algorithms under two parameter settings:

\textit{Baseline Parameter}: We adopt \(\alpha=150, \beta=10^{-6}\) for \(\text{2R-Large}_{\bigtriangleup}\) and \(\text{2R-Small}_{\bigtriangleup}\), following Imola et al.~\cite{imola2022communication}. For our algorithms, we set \(\alpha=150\), \(\beta=10^{-3}\) to bound the relative bias by \(2\beta\) (\autoref{pro:clip}). \(\epsilon\) is allocated with ratio \(\epsilon_0:\epsilon_1:\epsilon_2=0.1:0.45:0.45\).

\textit{Our Parameter}: Based on our theoretical bias-variance trade-off (detailed in RQ2), we set \(\alpha=50\), \(\beta=10^{-2}\) for TriMTR, and \(\beta=10^{-1}\) for QuaTR. For fair comparison, baselines from~\cite{imola2022communication} are also run with \(\alpha=50, \beta=10^{-1}\).

\textbf{For RQ2.} We perform a parameter sweep over \(\alpha\) and \(\beta\) to empirically validate our theoretical bias-variance trade-off analysis. Based on this analysis, we identify and justify the optimal parameters used in the \textit{Our Parameter} setting above, providing practical deployment guidance.

\textbf{For RQ3.} We perform an ablation study: \textit{Step~1} applies only graph projection; \textit{Step~2} adds one-round noise; \textit{Step~3} adds clipping (omitted for TriTR/\(\text{TriTR}^*\)); \textit{Step~4} adds second-round noise.

\subsection{Experimental Results}

\begin{figure*}[t]
  \centering
  \includegraphics[width=\textwidth]{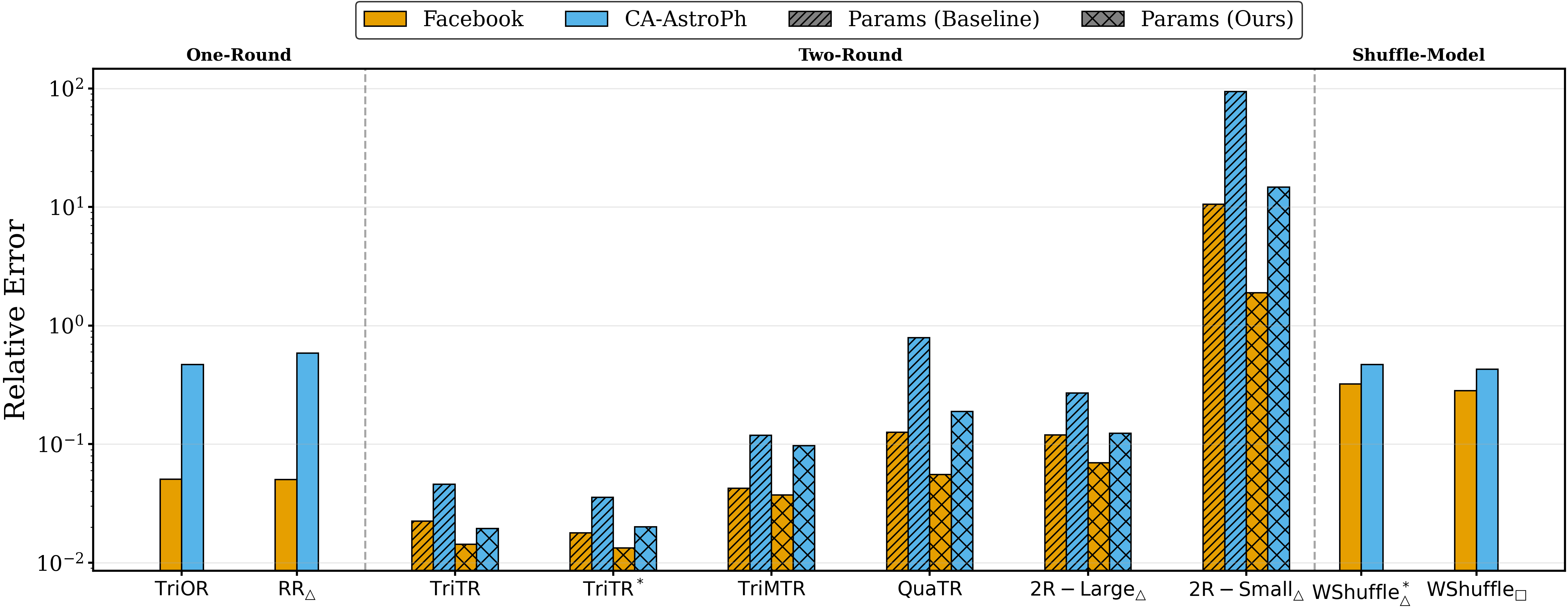}
  \caption{\textbf{Relative Error Comparison of Our Algorithms against Baseline Algorithms.}}
  \label{fig:Algorithm_Comparison}
\end{figure*}

\textbf{\autoref{fig:Algorithm_Comparison} answers RQ1.} For triangle counting, TriOR performs the same as \(\text{RR}_{\bigtriangleup}\), matching the proof in \autoref{app:Equ}. Our two-round algorithms consistently outperform all baselines under both settings. Under the \textit{Baseline Parameter}, \(\text{TriTR}^*\) beats TriTR (equivalent to TDP-VC~\cite{he2025robust}) and \(\text{2R-Large}_{\bigtriangleup}\), while TriMTR surpasses \(\text{2R-Small}_{\bigtriangleup}\) and \(\text{WShuffle}_{\bigtriangleup}^*\). Under the \textit{Our Parameter}, this superiority is further consolidated. 

For quadrangle counting, QuaTR beats \(\text{WShuffle}_{\Box}\) on Facebook under both settings. On CA-AstroPh, QuaTR is comparable under the \textit{Baseline Parameter}, but demonstrates a clear lead under the \textit{Our Parameter}, highlighting the importance of theoretical tuning on sparse graphs.

\textbf{TriTR Optimization.} \autoref{fig:Algorithm_Comparison} reveals an interesting phenomenon regarding bound tightness. Under the \textit{Baseline Parameter} ($\alpha=150$), \(\text{TriTR}^*\) significantly outperforms \(\text{TriTR}\) on both datasets. However, under the \textit{Our Parameter} ($\alpha=50$), \(\text{TriTR}^*\) loses its advantage on CA-AstroPh, where \(\text{TriTR}\) outperforms it (\autoref{tab:TriTR}). 

This reflects theoretical bound tightness: \(\text{TriTR}^*\) relies on \textit{Bernstein's inequality} which is not tight. The smaller \(\alpha\) restricts \(\tilde{d}_u\); for small \(\tilde{d}_u\), Bernstein's inequality becomes looser than the per-element bound (\(\kappa \tilde{d}_u < \kappa_u\)), causing \(\text{TriTR}^*\) to inject unnecessary noise. Conversely, for large \(d_u, \tilde{d}_u\) (e.g., Facebook dataset and the Baseline Parameter \(\alpha=150\)), \(\kappa_u\) becomes tighter than \(\kappa \tilde{d}_u\). To provide a robust solution, we recommend \(\text{TriTR}^2\), which adaptively selects the tightest bound \(\min(\kappa_u, \kappa \tilde{d}_u)\) to minimize noise and consistently match or surpass the best variant.

With the improved parameter settings, the concrete gains are as follows. We compare: (i) \(\text{TriTR}^2\) vs.\ TriTR; (ii) TriMTR vs.\ \(\text{2R-Small}_{\bigtriangleup}\) and \(\text{WShuffle}_{\bigtriangleup}^*\); and (iii) QuaTR vs.\ \(\text{WShuffle}_{\Box}\). 
On Facebook, \(\text{TriTR}^2\)'s RE is \(0.9287\times\) that of TriTR; TriMTR's RE is \(0.0198\times\) and \(0.1159\times\) that of \(\text{2R-Small}_{\bigtriangleup}\) and \(\text{WShuffle}_{\bigtriangleup}^*\), respectively; QuaTR's RE is \(0.1963\times\) that of \(\text{WShuffle}_{\Box}\). 
On CA-AstroPh, the corresponding ratios are \(0.9738\times\), \(0.0066\times\), \(0.2066\times\), and \(0.4407\times\).

\begin{table}[h]
  \centering
  \caption{\textbf{Comparison of Three TriTR Variants.}}
  \label{tab:TriTR}
  \begin{tabular}{|c|c|c|c|}
    \hline
    Datasets & \(\text{TriTR}\) & \(\text{TriTR}^*\) & \(\text{TriTR}^2\) \\
    \hline
    Facebook & \(1.43\times10^{-2}\) & \(1.33\times10^{-2}\) & \(1.33\times10^{-2}\) \\
    \hline
    CA-AstroPh & \(1.94\times10^{-2}\) & \(2.01\times10^{-2}\) & \(1.89\times10^{-2}\) \\
    \hline
  \end{tabular}
\end{table}

\begin{table}[htbp]
  \centering
  \caption{\textbf{Comparison with GenShuffle Algorithms.}}
  \label{tab:genshuffle}
  \begin{tabular}{|c|c|c|c|}
    \hline
    Subgraph & Algorithm & Facebook & CA-AstroPh \\
    \hline
    \multirow{4}{*}{Triangle} 
    & \(\text{GenShuffle}_1\) & \(1.05 \times 10^0\) & \(1.73 \times 10^0\) \\
    & \(\text{GenShuffle}_{20}\) & \(6.50 \times 10^{-1}\) & \(7.94 \times 10^0\) \\
    & $\text{WShuffle}_{\triangle}^*$ & \(3.23 \times 10^{-1}\) & \(4.69 \times 10^{-1}\) \\
    & \textbf{TriMTR} & \(\mathbf{3.74 \times 10^{-2}}\) & \(\mathbf{9.69 \times 10^{-2}}\) \\
    \hline
    \multirow{4}{*}{4-Cycle} 
    & \(\text{GenShuffle}_1\) & \(6.45\times 10^{-1}\) & \(3.58 \times 10^{-1}\) \\
    & \(\text{GenShuffle}_{20}\) & \(2.41 \times 10^{-1}\) & \(5.08 \times 10^{0}\) \\
    & $\text{WShuffle}_{\Box}$ & \(2.82 \times 10^{-1}\) & \(4.28 \times 10^{-1}\) \\
    & \textbf{QuaTR} & \(\mathbf{5.55\times 10^{-2}}\) & \(\mathbf{1.89 \times 10^{-1}}\) \\
    \hline
  \end{tabular}
\end{table}

\textbf{Comparison with GenShuffle.} We also compare against the recent GenShuffle framework~\cite{fang2025counting} and shuffle-based baselines~\cite{imola2022differentially}. As shown in \autoref{tab:genshuffle}, under the \textit{Our Parameter} setting, TriMTR and QuaTR consistently achieve the lowest relative errors. This confirms that our algorithms outperform the evaluated recent shuffle-model baselines under pure LDP without relying on a semi-trusted shuffler.

\textbf{Running time.} Given the variety of algorithms and their implementation-dependent efficiency, we report the running time only for the most critical one, \(\text{RR}_{\bigtriangleup}\) (\(O(n^3)\) complexity). In our experiments, it takes 0.0223 seconds per estimation on Facebook and 1.2651 seconds on CA-AstroPh. With GPU acceleration, this \(O(n^3)\) runtime is no longer prohibitive.

\begin{figure}[h]
  \centering
  \includegraphics[width=\linewidth]{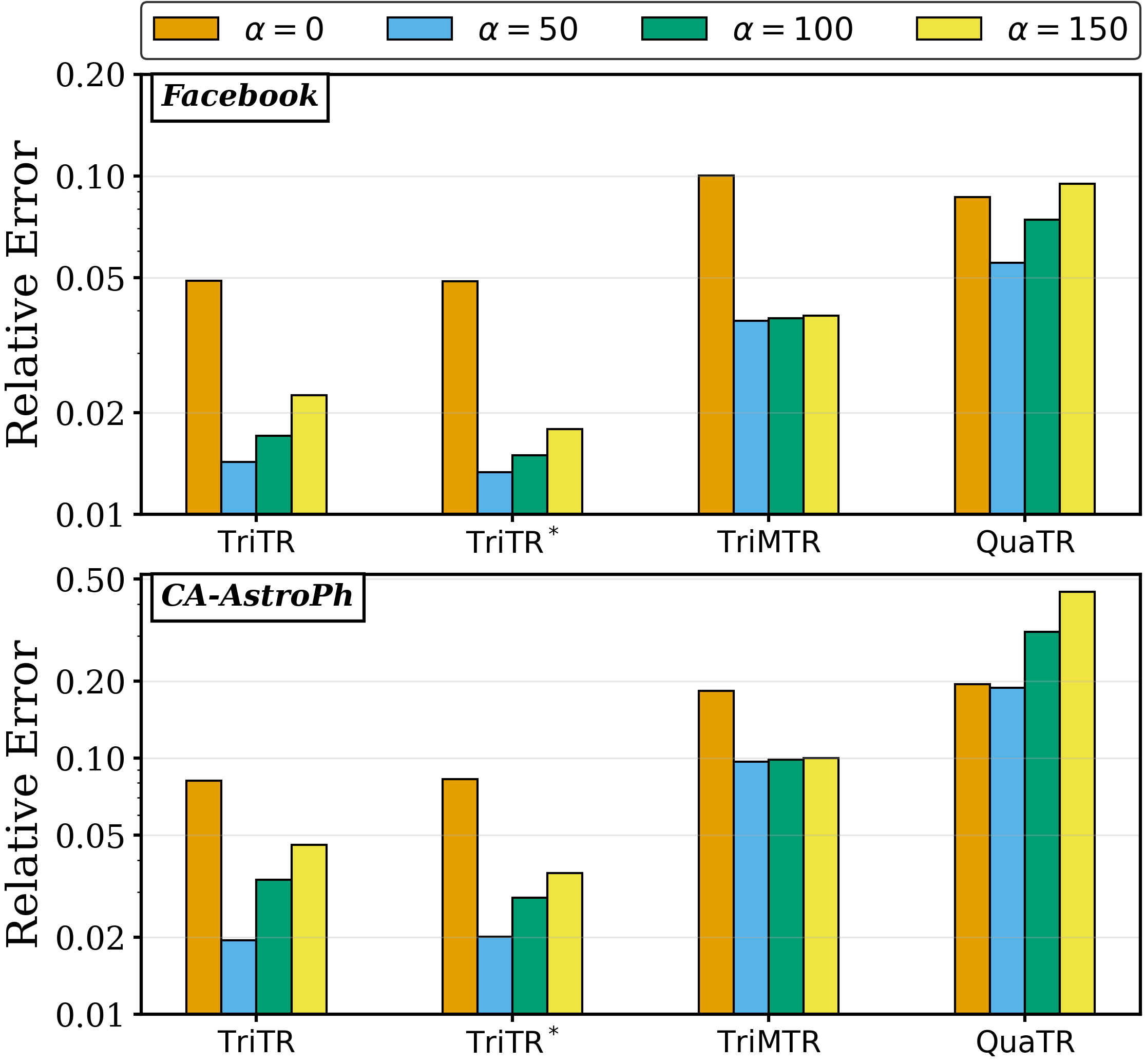}
  \caption{\textbf{Relative Error under Varying Graph Projection Parameter $\alpha$.}}
  \label{Alpha}
\end{figure}

\begin{figure}[h]
  \centering
  \includegraphics[width=\linewidth]{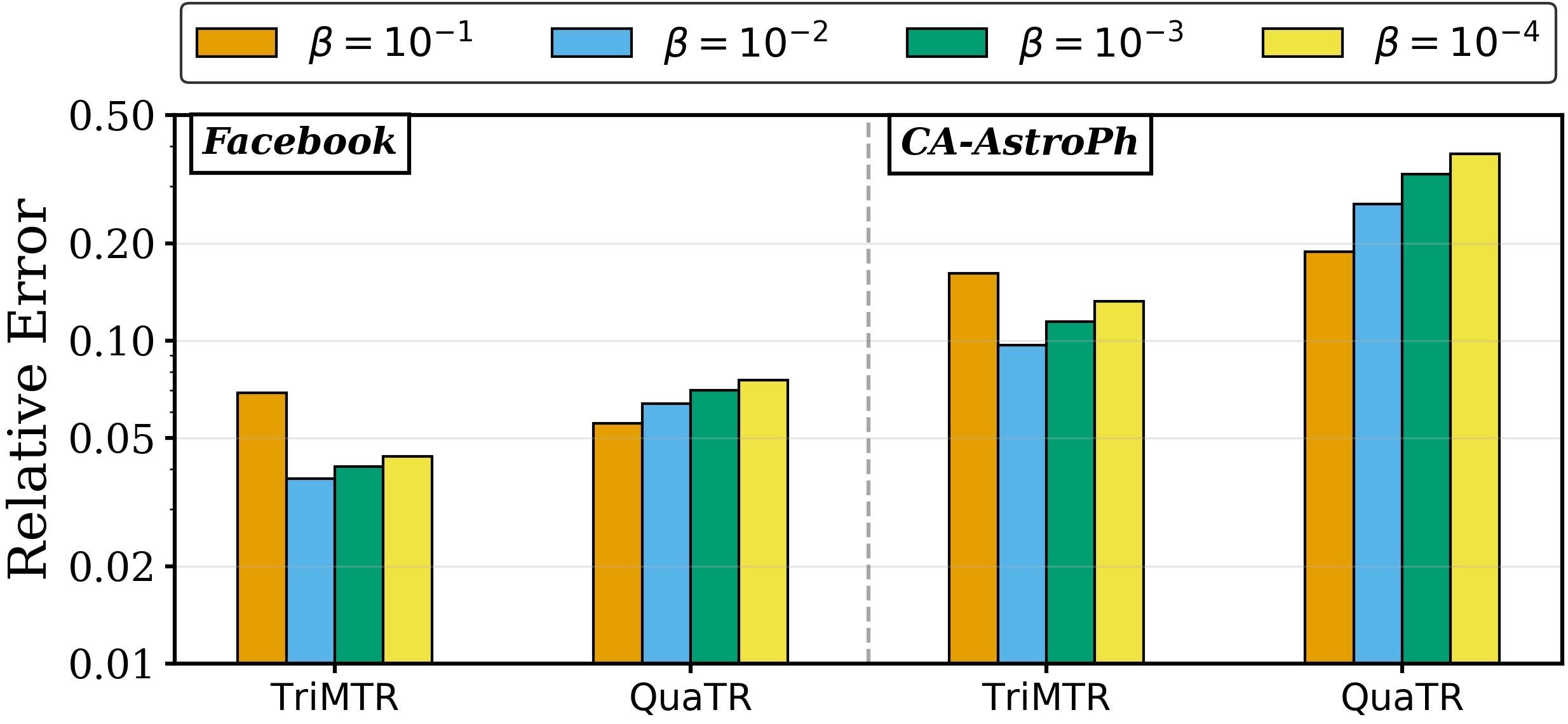}
  \caption{\textbf{Relative Error under Varying Clipping Parameter $\beta$.}}
  \label{Beta}
\end{figure}

\begin{figure}[h]
  \centering
  \includegraphics[width=\linewidth]{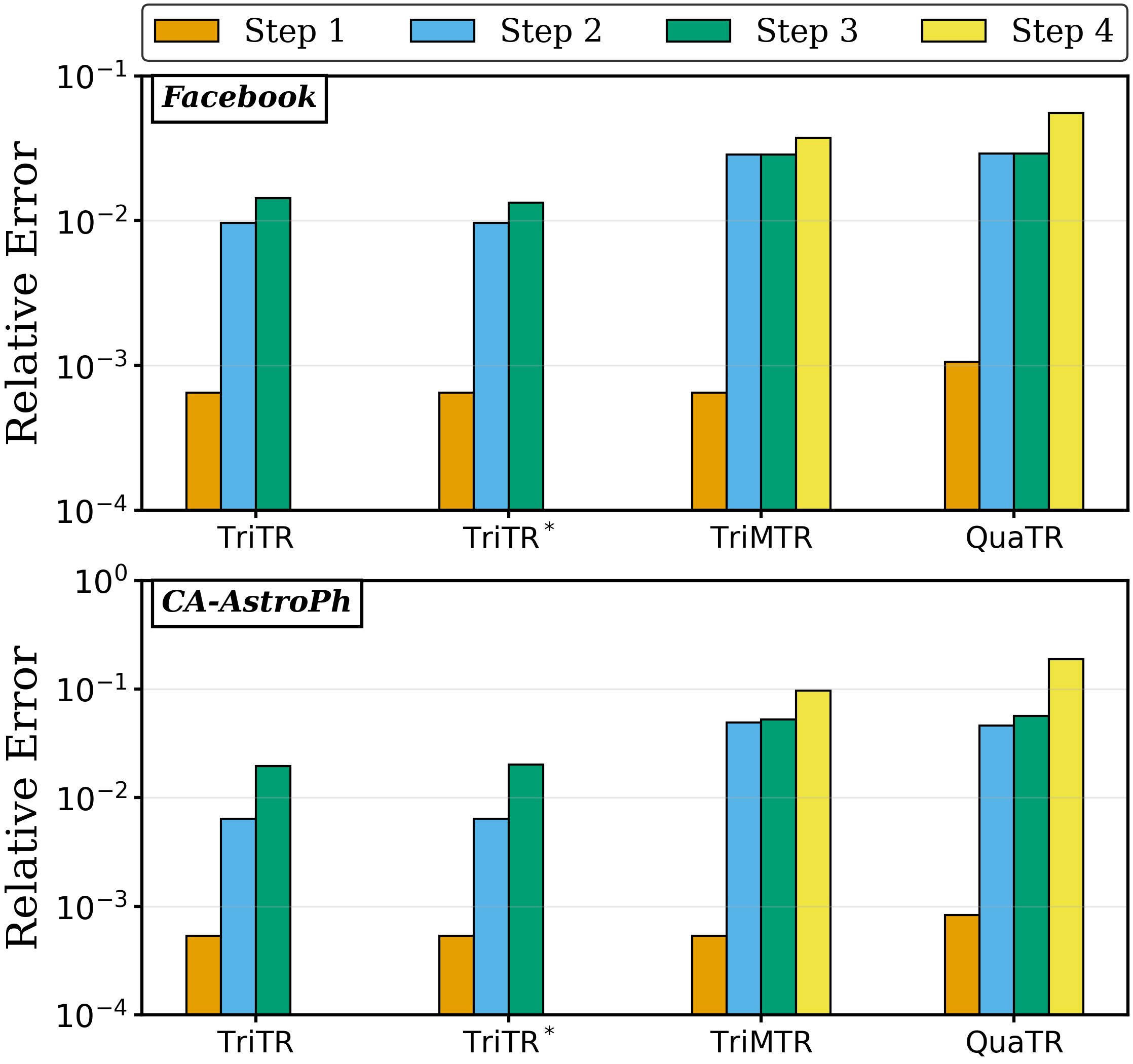}
  \caption{\textbf{Decomposition of Error Sources.}}
  \label{combined_step}
\end{figure}

\textbf{Deployment Optimization.} \textit{To reduce communication}, users can download \(\tilde{A}\) (1-bit per entry) instead of float \(\hat{A}\), as both convey the same information. \textit{To speed up computation}, \(\hat{B}\) can be computed via bit additions over \(\tilde{A}\), since \(\hat{A}\)'s entries yield only three product types—counting occurrences is faster than floating-point multiplications. Since \(\text{RR}_{\bigtriangleup}\) is equivalent to TriOR, we recommend deploying \(\text{RR}_{\bigtriangleup}\)~\cite{imola2021locally} for its single-bit operation efficiency.

\textbf{\autoref{Alpha}-\ref{Beta} answer RQ2.} All four two-round algorithms peak at \(\alpha=50\); TriMTR and QuaTR have optimal \(\beta=0.01\) and \(0.1\), respectively, far from baseline parameters \(\alpha=150,\beta=10^{-3}\). 

A larger \(\alpha\) reduces edge-removal probability and raises \(\tilde{d}_u\), yielding a looser \(\kappa\) and less bias, but increases second-round noise due to the larger \(\kappa\). At \(\alpha=50\), the edge-removal probability is \(3.37\times10^{-3}\); further increasing \(\alpha\) offers less bias reduction than the resulting variance increase.

A smaller \(\beta\) reduces clipping bias but increases \(\kappa\) and second-round noise. Clipping also shrinks one-round noise variance (e.g., for \(X\sim\mathcal{N}(\mu,\sigma^2)\), \(\mathbb{V}(\operatorname{clip}(X,\kappa))<\sigma^2\)). The optimal \(\beta\) differs because TriMTR's \(\kappa_u\) is tighter than QuaTR's \(\kappa\), making it more prone to clipping and thus requiring a smaller \(\beta\).

\textbf{\autoref{combined_step} answers RQ3.} Using the optimal parameters from RQ2, adjacent bar differences reflect each component's effect. Bias from graph projection and clipping is negligible compared to one‑ and two‑round noise. Though Step~1 appears high, the y‑axis is logarithmic, so its actual bias is less than \(0.1\times\) the one‑round noise error.

Although \autoref{pro:clip} bounds clipping bias by \(2\beta\), practical bias is far smaller: the maximum relative err in \autoref{pro:clip} is approached as \(\mu\to0\), which is rare; \(\alpha>0\) provides slack and reduces clipping probability; and \(\kappa_u,\kappa \tilde{d}_u\) are derived in worst‑case graph structure, while real graphs have smaller expectation and variance, further lowering clipping.

\section{Conclusion}

We introduced the NAM and DPRD frameworks for private subgraph counting under edge-LDP without trusted third parties. These frameworks led to a series of triangle and quadrangle counting algorithms, including TriOR, TriTR, TriMTR, QuaTR, and \(\text{TriTR}^*\). Our analysis provided the first exact closed-form MSE for the one-round triangle-counting class and the first relative-error upper bound showing stable accuracy as graph size grows under stable clustering coefficient and average degree. Experiments confirmed that our algorithms achieve higher accuracy than the evaluated baselines while maintaining strict edge-LDP guarantees. Moving forward, we plan to extend the DPRD framework to a broader range of privacy-preserving learning scenarios.

\cleardoublepage
\appendix
\section*{Ethical Considerations}
\textbf{Stakeholders:} 
The primary stakeholders in this research include the users of social networks (data owners), data collectors (e.g., service providers and major technology companies), and potential adversaries.

\textbf{Impacts:} 
The core problem addressed in this paper is the preservation of users' edge privacy (such as friend lists in social networks) under the assumption of an untrusted data collector. In this model, users do not trust the collector (e.g., large tech platforms) to keep their raw data confidential after upload. The potential impact involves a trade-off: data collectors require accurate subgraph counting to analyze network structures for commercial optimization, while users face the risk of privacy leakage. A malicious adversary—whether it be the untrusted collector itself or an external attacker intercepting the transmission—aims to compromise this edge privacy.

\textbf{Mitigations: }To mitigate these privacy risks, we adopt the Edge Local Differential Privacy (Edge-LDP) framework. A core advantage of this approach is its robustness against a malicious data collector. By ensuring that noise is added locally on the user’s device before any data leaves their control, the collector never accesses the raw edge data. Consequently, even if the collector acts maliciously, colludes with other parties, or its internal servers are compromised, the users’ actual private edge information remains mathematically protected against leakage. Furthermore, compared to prior work, our proposed method provides more accurate subgraph counting estimates under the same privacy budget, thereby maximizing the data utility and commercial value for the data collector under the stated edge-LDP guarantees and parameter choices. In our two-round protocols, the noisy matrices downloaded by users in the second round are already DP-protected outputs. According to the post-processing immunity theorem of differential privacy, any computation or download of these noisy matrices cannot weaken the established privacy guarantee, thus preventing additional edge inference risks.

\textbf{Decision: }We believe the publication of this work is justified as it improves the utility/privacy trade-off under the stated threat model. Our approach provides formal $(\epsilon, \delta)$-edge LDP guarantee against edge inference attacks while enabling data collectors to perform necessary network optimizations through more precise statistical analysis. This creates a beneficial outcome where commercial utility is enhanced while upholding rigorous privacy standards for users.

\cleardoublepage

\section*{Open Science}

To facilitate the reproducibility of our results and support further research, we have made all research artifacts used in this paper publicly available.

\textbf{Artifact Availability:} 
All datasets utilized in this paper, the complete source code for our experiments (including the implementations of both our proposed algorithms and the benchmark algorithms used for comparison), and all experimental results are open-sourced in an anonymized repository.

\textbf{Access:} 
Reviewers can access the anonymous GitHub repository via the following link: 
\url{https://anonymous.4open.science/r/Locally_Private_Subgraph_Counting_via_NAM_DPRD-A421/}

\cleardoublepage
\bibliographystyle{plain}
\bibliography{RelatedWork}

\cleardoublepage
\appendix

\begin{table*}[t]
\centering
\caption{Priority rules for triangle counting. The relative ordering of Rules 4 and 5 depends on graph sparsity (see discussion).}
\label{tab:triangle-deployment-guide}
\begin{tabular}{p{0.08\linewidth}p{0.58\linewidth}p{0.28\linewidth}}
\toprule
Priority & Deployment regime & Recommended method \\
\midrule

1
&
Multi-round interaction is acceptable, and users can tolerate an
$O(n^2)$-scale noisy-graph download.
&
$\mathrm{TriTR}^*$ if $(\epsilon,\delta)$-DP is
acceptable; otherwise $\mathrm{TriTR}$, TDP-VC~\cite{he2025robust}.
\\
\midrule

2
&
Multi-round interaction is acceptable, users cannot tolerate an
$O(n^2)$-scale download, and the analyst can afford $O(n^3)$
computation.
&
TriMTR
\\
\midrule

3
&
Multi-round interaction is not acceptable, the analyst cannot afford
$O(n^3)$ computation, and users accept the shuffle model.
&
$\text{WShuffle}_{\triangle}^*$~\cite{imola2022differentially}, $\mathrm{GenShuffle}_{\triangle}$~\cite{fang2025counting}.
\\
\midrule

4 or 5
&
Multi-round interaction is not acceptable, and the analyst can afford
$O(n^3)$ computation.
&
TriOR, $\mathrm{RR}_{\triangle}$~\cite{imola2021locally}, 
TCRR~\cite{eden2025triangle}.
\\
\midrule

4 or 5
&
Multi-round interaction is acceptable, users cannot tolerate an
$O(n^2)$-scale download and the analyst cannot afford $O(n^3)$
computation.
&
$\mathrm{2R\mbox{-}Small}_{\triangle}$~\cite{imola2022communication}
\\
\midrule

6
&
Multi-round interaction is not acceptable, the analyst cannot afford
$O(n^3)$ computation, and users do not accept the shuffle model.
&
$\mathrm{WLocal}_{\triangle}$~\cite{imola2022differentially}
\\

\bottomrule
\end{tabular}
\end{table*}

\begin{table*}[t]
\centering
\caption{Priority rules for quadrangle counting.}
\label{tab:quadrangle-deployment-guide}
\begin{tabular}{p{0.08\linewidth}p{0.58\linewidth}p{0.28\linewidth}}
\toprule
Priority & Deployment regime & Recommended method \\
\midrule

1
&
Multi-round interaction is acceptable, users can tolerate an
$O(n^2)$-scale noisy-graph download, and the analyst can afford the
\(O(n^3)\) computation.
&
$\mathrm{QuaTR}$.
\\
\midrule

2
&
Multi-round interaction is not acceptable, and users accept the
shuffle model.
&
\(\text{WShuffle}_{\Box}\)~\cite{imola2022differentially}, $\mathrm{GenShuffle}_{\Box}$~\cite{fang2025counting}.
\\
\midrule

3
&
Multi-round interaction is not acceptable, and users do not accept the
shuffle model.
&
$\mathrm{WLocal}_{\Box}$~\cite{imola2022differentially}
\\

\bottomrule
\end{tabular}
\end{table*}

\section{Practical Deployment Guidance}
\label{app:deployment-guidance}

Private subgraph counting algorithms should not be compared only by their asymptotic error bounds. The right choice also depends on the trust model, interaction rounds, user download cost, and analyst computation.

The following tables give priority-based selection rules. Choose the first applicable rule from top to bottom for your scenario. \autoref{tab:triangle-deployment-guide} covers triangle counting. \autoref{tab:quadrangle-deployment-guide} covers quadrangle counting. Note that Rules 4 and 5 in \autoref{tab:triangle-deployment-guide} are not strictly ordered. Specifically, $\mathrm{2R\mbox{-}Small}_{\triangle}$ performs better on sparse graphs, while TriOR performs better on dense graphs.

\section{Weighted Graphs and Differential Privacy}
\label{app:weighted}

\subsection{Weighted Graphs in Financial Flow Networks}
\label{app:weighted_motivation}

In many real-world systems, edges are naturally associated with real-valued intensities rather than binary presence indicators. A typical example is a financial transaction network, where companies are represented as nodes and cash flows between companies as directed edges. Suppose company \(i\) transfers a fraction \(a_{ij}\in[0,1]\) of its total outflow to company \(j\). Then a directed cycle \(i\to j\to k\to i\) represents a circular flow of capital, and the product
\[
a_{ij}\,a_{jk}\,a_{ki}
\]
measures the intensity of this three-party closed flow. Such quantities are meaningful for anti-money-laundering and financial risk monitoring, since large products indicate concentrated circular capital circulation.

Motivated by this, we extend our framework to weighted directed graphs. Let \(\mathcal{G}_{\mathrm{W}}\) denote the set of all weighted directed graphs without self-loops. We represent a weighted directed graph by a weight matrix
\[
A=(a_{ij})\in[0,1]^{n\times n},
\]
where \(a_{ij}\) is the weight of the directed edge from \(v_i\) to \(v_j\), and \(a_{ii}=0\) for all \(i\in[n]\). The condition \(a_{ij}=0\) means no edge from \(v_i\) to \(v_j\), while \(a_{ij}>0\) means a directed edge with weight \(a_{ij}\). Since the matrix is not required to be symmetric, this representation naturally covers directed financial flows.

For general nonnegative weights \(w_{ij}\), one may normalize them by a global public quantity, e.g.,
\[
a_{ij}=\frac{w_{ij}}{\max_{k\neq l} w_{kl}},
\]
so that all normalized weights lie in \([0,1]\). Since the normalization uses only public aggregate information, it does not introduce additional privacy leakage. If the original values are already proportions, as in the financial flow example above, then they are directly in \([0,1]\) without further transformation.

The unweighted undirected graphs studied in the main text can be viewed as the special case where every edge is undirected and has weight \(1\), i.e., \(a_{ij}=a_{ji}\in\{0,1\}\).

\subsection{Differential Privacy on Weighted Directed Graphs}
\label{app:weighted_dp}

We now define edge differential privacy for weighted directed graphs. Let the \(i\)-th row of the weight matrix \(A\),
\[
\mathbf{a}_i=(a_{i1},a_{i2},\ldots,a_{in})\in[0,1]^n,
\]
be the local neighbor list of user \(v_i\), representing the weights of its outgoing edges.

\begin{definition}[\((\epsilon,\delta)\)-edge LDP on weighted graphs]
\label{def:weighted_ldp}
Let \(\epsilon\in\mathbb{R}_{\ge 0}\) and \(\delta\in[0,1]\). A local randomizer \(\mathcal{R}\) with domain \([0,1]^n\) provides \((\epsilon,\delta)\)-edge LDP if for any two neighbor lists \(\mathbf{a}_i,\mathbf{a}_i'\in[0,1]^n\) that differ in at most one coordinate, and for any measurable \(S\subseteq \mathrm{Range}(\mathcal{R})\),
\[
\Pr[\mathcal{R}(\mathbf{a}_i)\in S]
\leq
e^{\epsilon}
\Pr[\mathcal{R}(\mathbf{a}_i')\in S]
+\delta.
\]
If \(\delta=0\), we say \(\mathcal{R}\) provides \(\epsilon\)-edge LDP.
\end{definition}

In this definition, two neighbor lists differing in one coordinate capture both changes in edge existence and changes in edge weight, because a coordinate can vary between \(0\) and any value in \([0,1]\).

\begin{definition}[\((\epsilon,\delta)\)-edge DP on weighted graphs]
\label{def:weighted_dp}
Let \(\epsilon\in\mathbb{R}_{\ge 0}\) and \(\delta\in[0,1]\). A randomized algorithm \(\mathcal{M}\) with domain \(\mathcal{G}_{\mathrm{W}}\) provides \((\epsilon,\delta)\)-edge DP if for any two weighted directed graphs \(G,G'\in\mathcal{G}_{\mathrm{W}}\) that differ in the weight of at most one directed edge, and for any measurable \(S\subseteq \mathrm{Range}(\mathcal{M})\),
\[
\Pr[\mathcal{M}(G)\in S]
\leq
e^{\epsilon}
\Pr[\mathcal{M}(G')\in S]
+\delta.
\]
If \(\delta=0\), we say \(\mathcal{M}\) provides \(\epsilon\)-edge DP.
\end{definition}

The neighborhood relation in \autoref{def:weighted_dp} means that there exists at most one ordered pair \((i,j)\), \(i\neq j\), such that \(a_{ij}\neq a'_{ij}\), and all other entries coincide. This generalizes the binary edge DP notion to weighted directed edges.

\textbf{Relation to the unweighted definitions.} When weights are restricted to \(\{0,1\}\), the above definitions reduce to the standard edge LDP and edge DP definitions for unweighted graphs used in the main text. The Laplace-based version of GNAM directly applies to this weighted setting, because the Laplace mechanism accepts real-valued inputs and preserves privacy for the same sensitivity bounds after projection to \([0,1]\). The NAM and DPRD frameworks therefore extend to weighted directed subgraph counting with only minor modifications to the local randomizer and aggregation procedure.

\section{Equivalence of Some One‑Round Algorithms}
\label{app:Equ}

\subsection{TriOR and TCRR}
We first theoretically demonstrate that TriOR with RR is mathematically equivalent to Triangle Counting via Randomized Response (TCRR)~\cite{eden2025triangle}. In TCRR, the variable $Y_{ij}$ is equivalent to the unbiased estimator $\hat{a}_{ij}$ in our context. The TCRR estimator can be reformulated as:

$$\sum_{i<j<k }^n Y_{ij}Y_{jk}Y_{ki} = \sum_{i = 1}^n \sum_{j = 1}^i \sum_{k = 1}^j \hat{a}_{ij} \hat{a}_{jk} \hat{a}_{ki} = \frac{1}{6} \text{tr}(\hat{A}^3)$$

This equivalence implies that TriOR inherits the theoretical optimality established for TCRR. To contextualize this, we restate the fundamental limits of non-interactive edge-LDP triangle counting as proven by Eden et al.~\cite{eden2025triangle}:

\begin{theorem}[Lower Bound \cite{eden2025triangle}]
For any $\epsilon > 0$, any one-round $\epsilon$-edge-LDP algorithm for triangle counting must incur an expected additive error of $\Omega(n^2)$.
\end{theorem}

\begin{theorem}[MSE of TCRR \cite{eden2025triangle}] \label{thm:eden_upper}
For any $\epsilon > 0$, the TCRR provides an unbiased estimator with variance:
\begin{equation}
\mathbb{V}(\hat{f}^{\bigtriangleup}_{\text{TCRR}}(G)) = \Theta\left(\frac{C_4(G)}{\epsilon^2} + \frac{n^{3}}{\epsilon^6}\right)
\end{equation}
where $C_4(G)$ denotes the number of quadrangles (4-cycles) in the graph $G$.
\end{theorem}

\autoref{thm:eden_upper} demonstrates that for graphs where $C_4(G) \leq \Theta(n^4)$, the additive error (square root of the variance) is $O(n^2)$, matching the lower bound and thus proving the asymptotic optimality of the RR-based approach. 

The significance of TriOR is twofold. First, by establishing the equivalence between TriOR and TCRR, we confirm that TriOR also achieves this theoretical lower bound. Second, while Eden et al. focus on asymptotic order-optimality (using $\Theta$ and $O$ notation), our \autoref{TriOR} provides the first exact \textbf{closed-form expression} for the MSE.

\subsection{TriOR and $\text{RR}_{\bigtriangleup}$}

We begin by establishing \autoref{lem:triple}.

\begin{lemma}\label{lem:triple}
Let $\tilde a_{ij}\in\{0,1\}$ be the binary randomized response output for edge $\{i,j\}$, and define
\[
\eta=e^{\epsilon},\qquad
\hat a_{ij}=\frac{(\eta+1)\tilde a_{ij}-1}{\eta-1}.
\]
For any unordered triple $\{i,j,k\}$, let $x=\tilde a_{ij}$, $y=\tilde a_{jk}$, $z=\tilde a_{ki}\in\{0,1\}$ and denote $r=x+y+z$. Then
\[
\hat a_{ij}\hat a_{jk}\hat a_{ki}
=\frac{(-1)^{\,3-r}\eta^{r}}{(\eta-1)^3}.
\]
\end{lemma}

\begin{proof}
For any $u\in\{0,1\}$, we have $(\eta+1)u-1 = (-1)^{1-u}\eta^{u}$. Hence
\[
\hat a_{ij}
=\frac{(\eta+1)\tilde a_{ij}-1}{\eta-1}
=\frac{(-1)^{1-\tilde a_{ij}}\eta^{\tilde a_{ij}}}{\eta-1}.
\]
Taking the product over $x,y,z$ and collecting exponents yields the claim.
\end{proof}

Then We can prove \autoref{thm:equiv}.

\begin{theorem}\label{thm:equiv}
Let $\mathcal T=\{\{i,j,k\}:1\le i<j<k\le n\}$ and, for $r\in\{0,1,2,3\}$, let
\[
m_r=\bigl|\{\{i,j,k\}\in\mathcal T:\ \tilde a_{ij}+\tilde a_{jk}+\tilde a_{ki}=r\}\bigr|.
\]
Define the two estimators
\[
\begin{aligned}
&\hat{f}^{\bigtriangleup}_{\text{RR}_{\bigtriangleup}}(G)
=\frac{1}{(\eta-1)^3}\bigl(\eta^3 m_3-\eta^2 m_2+\eta m_1-m_0\bigr)
\\
&\hat{f}^{\bigtriangleup}_{\text{TriOR}}(G)=\frac{1}{6}\operatorname{tr}(\hat A^3)=\sum_{i<j<k}\hat a_{ij}\hat a_{jk}\hat a_{ki}
\end{aligned}
\]
where $\hat A=(\hat a_{ij})$. Then for every fixed noisy matrix $\tilde A$,
\[
\hat{f}^{\bigtriangleup}_{\text{RR}_{\bigtriangleup}}(G)=\hat{f}^{\bigtriangleup}_{TriOR}(G).
\]
\end{theorem}

\begin{proof}
Grouping the sum $\sum_{i<j<k}\hat a_{ij}\hat a_{jk}\hat a_{ki}$ by the number $r$ of ones in each triple and applying \autoref{lem:triple} yields
\[
\begin{aligned}
\hat{f}^{\bigtriangleup}_{\text{TriOR}}(G))
&=\sum_{r=0}^3 m_r\frac{(-1)^{3-r}\eta^r}{(\eta-1)^3}\\
&=\frac{1}{(\eta-1)^3}\bigl(\eta^3 m_3-\eta^2 m_2+\eta m_1-m_0\bigr)\\
&=\hat{f}^{\bigtriangleup}_{\text{RR}_{\bigtriangleup}}(G)
\end{aligned}
\]
\end{proof}

\autoref{thm:arr_mse} characterizes the MSE of $\text{ARR}_{\triangle}$, which converges to $\text{RR}_{\triangle}$ under the parameterization $\mu = \frac{e^{\epsilon}}{e^{\epsilon} + 1}$. While the existing literature establishes a loose upper bound of $O(n^4)$ for the theoretical MSE of $\text{RR}_{\triangle}$, by demonstrating the equivalence between TriOR and $\text{RR}_{\triangle}$, we derive a tight closed-form expression for the latter, as denoted in \autoref{TriOR}.

\begin{theorem}[MSE of $\text{ARR}_{\triangle}$ \cite{imola2022differentially}] \label{thm:arr_mse}
When we treat the privacy budget $\epsilon$ as a constant, the $ARR_{\triangle}$ algorithm provides the following utility guarantee for the triangle count estimate $\hat{f}^{\triangle}(G)$:
\begin{equation}
\text{MSE}(\hat{f}^{\triangle}(G)) = \mathbb{V}(\hat{f}^{\triangle}(G)) = O\left(\frac{n^4}{\mu^6}\right)
\end{equation}
where $\mu \in [0, \frac{e^{\epsilon}}{e^{\epsilon}+1}]$ is a parameter of the Asymmetric Randomized Response (ARR).
\end{theorem}

\section{Computational Cost Estimation}
\label{sec:D}

We further discuss the practical feasibility of the collector-side
matrix multiplication involved in the NAM framework, especially the
computation of noisy matrix powers such as
\[
    \widehat{B}=\widehat{A}^{2}.
\]
Such matrix operations are performed entirely by the data collector and
do not increase the local computational burden of individual users. Users
only need to perform local randomization, upload their perturbed data, and
execute lightweight local operations in interactive protocols. Therefore,
the cost of matrix multiplication should be interpreted as an offline
collector-side analytics cost rather than a user-side deployment bottleneck.

To estimate the scale of this computation, consider a large graph with
\(n=10^{6}\). Using standard dense matrix multiplication, one
\(n\times n\) matrix multiplication requires approximately
\[
    2n^{3}=2.0\times 10^{18}
\]
floating-point operations. Although this number appears large, it is not
prohibitive from the perspective of modern GPU computation. For example,
the NVIDIA GeForce RTX 5090, a high-end consumer GPU, has a peak FP32
throughput of approximately \(104.8\) TFLOPS and \(32\)GB of GDDR7 memory.
Using \(104.8\times 10^{12}\) FLOP/s as an ideal throughput estimate, the
arithmetic time is approximately
\[
    \frac{2.0\times 10^{18}}{104.8\times 10^{12}}
    \approx 1.91\times 10^{4}\ \mathrm{seconds}
    \approx 5.3\ \mathrm{hours}.
\]
In practice, the actual running time depends on sustained GPU utilization,
I/O bandwidth, data reuse, and implementation quality. Nevertheless, even
if the sustained throughput is only a fraction of the peak throughput, the
computation is typically on the order of hours to one day. For periodic
large-scale graph analytics, such collector-side offline computation is
acceptable, especially since collecting, randomizing, uploading, and
scheduling data from a large population of users may already take a
substantial amount of time.

Regarding memory, a common misconception is that computing
\(\widehat{A}^{2}\) requires all dense matrices to reside in GPU memory
simultaneously. In fact, the static storage size of the full matrices and
the peak working memory during computation are different concepts. In FP32,
one \(10^{6}\times 10^{6}\) dense matrix requires
\[
    10^{12}\times 4\ \mathrm{bytes}=4\ \mathrm{TB}
\]
of storage. If two input matrices and one output matrix are materialized,
the total static storage is approximately \(12\)TB. This size is indeed far
beyond the memory capacity of a single GPU, but it mainly reflects the
storage size of the full matrices rather than the unavoidable peak GPU
memory requirement. For a data-analysis server or workstation, terabyte-scale
static storage can be handled by SSDs, NVMe storage, external storage, or
distributed storage systems.

The actual computation can be implemented using standard blocked or
out-of-core matrix multiplication. For example, let the block size be
\(b=10000\). Since
\[
    n/b = 10^{6}/10000 = 100,
\]
the output matrix can be partitioned into \(100\times 100=10000\) output
blocks. For an output block
\(\widehat{B}_{I,J}\in\mathbb{R}^{10000\times 10000}\), a simple blocked
implementation only needs to load
\[
    \widehat{A}_{I,:}\in\mathbb{R}^{10000\times 10^{6}}
    \quad\text{and}\quad
    \widehat{A}_{:,J}\in\mathbb{R}^{10^{6}\times 10000}.
\]
In FP32, these two blocks require approximately \(40\)GB each, or
approximately \(80\)GB in total, plus about \(0.4\)GB for the output block
\(\widehat{B}_{I,J}\). Thus, even this simple two-dimensional blocking
strategy reduces the peak working memory from the terabyte scale to roughly
the hundred-gigabyte scale. If a lower memory footprint is desired, one can
use a smaller block size, such as \(5000\times 5000\), or further block the
middle dimension. Such blocking strategies do not change the total
\(O(n^{3})\) arithmetic complexity, but they substantially reduce the peak
working memory. Moreover, dense matrix multiplication is a standard and
highly optimized workload for GPUs and GPU servers.

Therefore, the \(O(n^{3})\) matrix multiplication involved in the NAM
framework should not be regarded as an inherent obstacle to deployment.
Rather, the main cost is shifted to the data collector, where mature
engineering solutions---including GPU acceleration, blocked computation,
out-of-core execution, and server-side storage---can be directly applied.
For offline global graph analytics, this computational cost is manageable:
it is not only feasible for hyperscale service providers, but also within
reach for organizations equipped with high-end workstations or small
server-level infrastructure.

\section{Proofs of Statements}

\subsection{Proof of \autoref{thm:1}}
\label{proof:thm:1}

\begin{proof}
By \autoref{def:4}, the noisy adjacency matrix $\hat{A}$ can be represented as $\hat{A} = A + X$, where $A$ is the original adjacency matrix and $X$ is a symmetric noise matrix satisfying:
\begin{enumerate}
    \item $\mathbb{E}[X] = 0$;
    \item $X^\top = X$ and $x_{ii} = 0$ for all $i \in [n]$;
    \item $x_{ij} \perp x_{kl}$ for any distinct edge sets $\{i,j\} \neq \{k,l\}$.
\end{enumerate}
We now prove the unbiasedness properties.

\paragraph{Property 1 (Unbiasedness of $\hat{B}$)} 

Expanding $\hat{B} = (A + X)^2$ yields $\hat{B} = A^2 + AX + XA + X^2$. For any $i \neq j$, the entry $\hat{b}_{ij}$ is:
\begin{equation*}
    \hat{b}_{ij} = b_{ij} + \sum_{k=1}^n a_{ik}x_{kj} + \sum_{k=1}^n x_{ik}a_{kj} + \sum_{k=1}^n x_{ik}x_{kj}.
\end{equation*}
Taking the expectation and applying linearity:

\textbf{First-order terms}: $\mathbb{E}[\sum a_{ik}x_{kj}] = \sum a_{ik} \mathbb{E}[x_{kj}] = 0$.

\textbf{Second-order terms}: For the term $\sum x_{ik}x_{kj}$, since $i \neq j$, the index pairs $\{i,k\}$ and $\{k,j\}$ represent distinct edges for all $k$ (if $k=i$ or $k=j$, the term is zero as $x_{ii}=0$). Thus, $x_{ik}$ and $x_{kj}$ are independent, giving $\mathbb{E}[x_{ik}x_{kj}] = \mathbb{E}[x_{ik}]\mathbb{E}[x_{kj}] = 0$.

Therefore, $\mathbb{E}[\hat{b}_{ij}] = b_{ij}$ for all $i \neq j$.

\paragraph{Property 2 (Unbiasedness of $\hat{C}$)} 

The expansion of $\hat{C} = (A + X)^3$ yields:
\begin{equation*}
\begin{aligned}
    \hat{C} = A^3 &+ (A^2X + AXA + XA^2) \\
    &+ (AX^2 + XAX + X^2A) + X^3.
\end{aligned}
\end{equation*}
For each diagonal entry $\hat{c}_{ii}$, we have:
\begin{equation}
\label{eq:c_ii_expansion}
\begin{aligned}
    \hat{c}_{ii} = c_{ii} &+ \sum_{j,k} (a_{ij}a_{jk}x_{ki} + a_{ij}x_{jk}a_{ki} + x_{ij}a_{jk}a_{ki}) \\
    &+ \sum_{j,k} (a_{ij}x_{jk}x_{ki} + x_{ij}a_{jk}x_{ki} + x_{ij}x_{jk}a_{ki}) \\
    &+ \sum_{j,k} x_{ij}x_{jk}x_{ki}.
\end{aligned}
\end{equation}
We evaluate the expectation of each block:

\textbf{First-order terms}: Since $\mathbb{E}[x] = 0$, the expectation is zero.

\textbf{Second-order terms}: For $a_{ij}x_{jk}x_{ki}$, the term is non-zero only if $\{j,k\} = \{k,i\}$, which implies $j=i$. However, $a_{ii}=0$ by definition. The same logic applies to other second-order terms.

\textbf{Third-order terms}: For $x_{ij}x_{jk}x_{ki}$, independence ensures the expectation is zero if all edges are distinct. If any indices are equal, the term vanishes because $x_{ii}=0$.

Consequently, $\mathbb{E}[\hat{c}_{ii}] = c_{ii}$ for all $i \in [n]$.
\end{proof}

\subsection{Proof of \autoref{pro:GNAM}}

\begin{proof}
Dwork et al. \cite{dwork2014algorithmic} proved that both the Laplace mechanism and Randomized Response preserve $(\epsilon, 0)$-DP, satisfying \autoref{def:1}. For any adjacency list $\mathbf{a}_i \in \{0, 1\}^n$, the sensitivity is $\Delta f = 1$ under the definition of $\epsilon$-edge LDP (\autoref{def:3}). Therefore, adding noise sampled from $\text{Lap}(1/\epsilon)$ preserves $\epsilon$-edge LDP. Regarding RR, it is straightforward to verify that \autoref{eq:4} satisfies \autoref{def:3}. Consequently, GNAM preserves $\epsilon$-edge LDP when employing either RR or the Laplace mechanism. By \autoref{pro:1}, it further follows that GNAM preserves $2\epsilon$-edge DP.
\end{proof}

\subsection{\autoref{lem:2} and Its Proof}

To facilitate subsequent proofs, let us derive \autoref{lem:2} at first.

\begin{lemma}
\label{lem:2}
Let $\{Z_i\}_{i \in \mathbb{N}}$ be a sequence of mutually independent random variables with $\mathbb{E}[Z_i] = 0$ for all $i$. Let $\mathcal{P}_1 = \prod_{i=1}^{n_1} Z_{\alpha_i}^{k_i}$ and $\mathcal{P}_2 = \prod_{j=1}^{n_2} Z_{\beta_j}^{l_j}$ be two products where indices $\alpha_i, \beta_j \in \mathbb{N}$ and exponents $k_i, l_j \in \{1, 2\}$. If $\mathbb{E}[\mathcal{P}_1] = \mathbb{E}[\mathcal{P}_2] = 0$, and there exists an index $m \in (\alpha \cup \beta) \setminus (\alpha \cap \beta)$ such that its corresponding exponent $q_m$ in the joint product $\mathcal{P}_1 \mathcal{P}_2$ is $1$, then:
\[
    \mathbb{V}(\mathcal{P}_1 + \mathcal{P}_2) = \mathbb{V}(\mathcal{P}_1) + \mathbb{V}(\mathcal{P}_2).
\]
\end{lemma}

\begin{proof}
By the definition of variance, we have:
\[
    \mathbb{V}(\mathcal{P}_1 + \mathcal{P}_2) = \mathbb{E}[(\mathcal{P}_1 + \mathcal{P}_2)^2] - (\mathbb{E}[\mathcal{P}_1 + \mathcal{P}_2])^2.
\]
Given the assumption $\mathbb{E}[\mathcal{P}_1] = \mathbb{E}[\mathcal{P}_2] = 0$, it follows that $\mathbb{E}[\mathcal{P}_1 + \mathcal{P}_2] = 0$. Expanding the quadratic term, we obtain:
\[
    \mathbb{V}(\mathcal{P}_1 + \mathcal{P}_2) = \mathbb{E}[\mathcal{P}_1^2] + \mathbb{E}[\mathcal{P}_2^2] + 2\mathbb{E}[\mathcal{P}_1 \mathcal{P}_2].
\]
Since $\mathbb{E}[\mathcal{P}_1] = 0$, we have $\mathbb{V}(\mathcal{P}_1) = \mathbb{E}[\mathcal{P}_1^2] - (\mathbb{E}[\mathcal{P}_1])^2 = \mathbb{E}[\mathcal{P}_1^2]$. Similarly, $\mathbb{V}(\mathcal{P}_2) = \mathbb{E}[\mathcal{P}_2^2]$. To prove the lemma, it suffices to show that $\mathbb{E}[\mathcal{P}_1 \mathcal{P}_2] = 0$.

Consider the product $\mathcal{P}_1 \mathcal{P}_2 = \prod_{h \in \alpha \cup \beta} Z_h^{q_h}$, where $q_h$ is the aggregate exponent for index $h$. By the lemma's condition, there exists $m \in (\alpha \cup \beta) \setminus (\alpha \cap \beta)$ such that $q_m = 1$. Due to the mutual independence of $\{Z_i\}$, the expectation of the product factors as:
\[
    \mathbb{E}[\mathcal{P}_1 \mathcal{P}_2] = \mathbb{E}[Z_m^{q_m}] \cdot \mathbb{E}\left[ \prod_{h \neq m} Z_h^{q_h} \right].
\]
As $q_m = 1$ and $\mathbb{E}[Z_m] = 0$, we have $\mathbb{E}[\mathcal{P}_1 \mathcal{P}_2] = 0$. Thus, the covariance term vanishes, yielding $\mathbb{V}(\mathcal{P}_1 + \mathcal{P}_2) = \mathbb{V}(\mathcal{P}_1) + \mathbb{V}(\mathcal{P}_2)$.
\end{proof}

\subsection{Proof of \autoref{TriOR}}

\begin{proof}
Following the notation in \autoref{proof:thm:1}, let $\hat{A} = A + X$. The unbiasedness of the estimator follows directly from the properties established in \autoref{thm:1}:
\[
    \mathbb{E} \left[ \hat{f}^{\triangle}(G) \right] = \mathbb{E} \left[ \frac{\text{tr}(\hat{A}^3)}{6} \right] = \frac{\text{tr}(A^3)}{6} = f^{\triangle}(G).
\]
Since the estimator is unbiased, the MSE is equivalent to the variance, i.e., $\text{MSE}(\hat{f}^{\triangle}(G)) = \mathbb{V}(\hat{f}^{\triangle}(G))$. Recall the trace expansion of $\hat{A}^3$:
\begin{equation*}
\begin{aligned}
    \text{tr}(\hat{A}^3) = \text{tr}(A^3) &+ 3\sum_{i,j,k} a_{ij}a_{jk}x_{ki} \\
    &+ 3\sum_{i,j,k} a_{ij}x_{jk}x_{ki} + \sum_{i,j,k} x_{ij}x_{jk}x_{ki}.
\end{aligned}
\end{equation*}
By \autoref{lem:2}, the cross-covariance terms vanish, allowing the total variance to be decomposed as:
\begin{equation*}
\begin{aligned}
    \mathbb{V}(\text{tr}(\hat{A}^3)) = & 9\mathbb{V}\left( \sum_{i,j,k} a_{ij}a_{jk}x_{ki} \right) \\
    & + 9\mathbb{V}\left( \sum_{i,j,k} a_{ij}x_{jk}x_{ki} \right) + \mathbb{V}\left( \sum_{i,j,k} x_{ij}x_{jk}x_{ki} \right).
\end{aligned}
\end{equation*}

We evaluate each variance component individually. For the first-order noise term, the sum simplifies to $2\sum_{i < j} b_{ij}x_{ij}$ in undirected graphs, where $b_{ij}$ denotes the number of length-2 paths between $v_i$ and $v_j$. With $\mathbb{V}(x_{ij}) = \sigma^2$, its variance contribution is:
\begin{equation}
\label{eq:var_first}
    9\mathbb{V}\left( \sum_{i,j,k} a_{ij}a_{jk}x_{ki} \right) = 36\sigma^2 \sum_{i < j} b_{ij}^2.
\end{equation}

For the second-order noise term, the expression counts structures where an existing edge $(v_i, v_j) \in E$ is completed by two noise edges $x_{jk}x_{ki}$. This simplifies to $2 \sum_{(v_i, v_j) \in E} \sum_{k \neq i,j} x_{ik}x_{kj}$. Since $\mathbb{V}(x_{ik}x_{kj}) = \sigma^4$, the variance contribution is:
\begin{equation}
\label{eq:var_second}
    9\mathbb{V}\left( \sum_{i,j,k} a_{ij}x_{jk}x_{ki} \right) = 36\sigma^4 (n-2)|E|.
\end{equation}

The third-order noise term represents noise-only triangles, which can be expressed as $6\sum_{i < j < k} x_{ij}x_{jk}x_{ki}$. Given $\mathbb{V}(x_{ij}x_{jk}x_{ki}) = \sigma^6$, its variance is:
\begin{equation}
\label{eq:var_third}
    \mathbb{V}\left( \sum_{i,j,k} x_{ij}x_{jk}x_{ki} \right) = 6\sigma^6 n(n-1)(n-2).
\end{equation}

Combining these results and dividing by $6^2$, we obtain the total MSE:
\begin{equation*}
\sigma^2 \sum_{i < j} b_{ij}^2 + \sigma^4 (n-2)|E| + \frac{1}{6} \sigma^6 n(n-1)(n-2).
\end{equation*}

It holds that $|E| \leq \frac{1}{2}nd_{\max}$ and $\sum_{i<j} b_{ij}^2 \le nd_{\max}^3$. Substituting these into MSE yields:
\[
    \text{MSE}(\hat{f}^{\triangle}(G)) \le \sigma^2 nd_{\max}^3 + \frac{1}{2}\sigma^4 n^2 d_{\max} + \frac{1}{6}\sigma^6 n^3,
\]
implying $\text{MSE}(\hat{f}^{\triangle}(G)) = O(nd_{\max}^3 + n^3)$.
\end{proof}

\subsection{Proof of \autoref{thm:4}}
\label{proof:thm:4}

\begin{proof}
Let $\hat{A} = A + X$ denote the noisy adjacency matrix, where $X$ satisfies the properties defined in Theorem 1. For each node $u \in [n]$, we define the local estimator $t_u$ as:
\begin{equation*}
\begin{aligned}
    t_u &= \sum_{(i,j): a_{ui}=a_{uj}=1} \hat{a}_{ij} = \sum_{i,j} a_{ui}a_{uj}(a_{ij} + x_{ij}) \\
    &= 2f_u^{\triangle} + \sum_{i,j} a_{ui}a_{uj}x_{ij},
\end{aligned}
\end{equation*}
where $f_u^{\triangle}$ is the number of triangles that include node $u$. Summing across all nodes, we obtain:
\begin{equation*}
    \sum_{u=1}^n t_u = 2\sum_{u=1}^n f_u^{\triangle} + \sum_{u,i,j} a_{ui}a_{uj}x_{ij}.
\end{equation*}
By the handshake lemma for triangles, $\sum_{u=1}^n f_u^{\triangle} = 3f^{\triangle}(G)$. For the noise term, utilizing the symmetry of $A$ and $X$, we rewrite the triple summation as:
\begin{equation*}
    \sum_{u,i,j} a_{ui}a_{uj}x_{ij} = \sum_{i,j,k} a_{ij}a_{jk}x_{ki}.
\end{equation*}
From the proof of \autoref{thm:1}, we have established that $\mathbb{E}[\sum_{i,j,k} a_{ij}a_{jk}x_{ki}] = 0$, which ensures the unbiasedness of the estimator:
\begin{equation*}
    \mathbb{E}\left[ \frac{1}{6} \sum_{u=1}^n t_u \right] = \frac{1}{6} \cdot 2 \cdot 3f^{\triangle}(G) = f^{\triangle}(G).
\end{equation*}

The variance of the global estimator is determined by the first-order noise component derived in \autoref{eq:var_first}. Specifically:
\begin{equation}
\label{eq:var_t_u}
\begin{aligned}
    \mathbb{V}\left( \frac{1}{6} \sum_{u=1}^n t_u \right) &= \frac{1}{36} \mathbb{V}\left( \sum_{i,j,k} a_{ij}a_{jk}x_{ki} \right) \\
    &= \frac{1}{36} \cdot 4\sigma^2 \sum_{i < j} b_{ij}^2 = \frac{1}{9}\sigma^2 \sum_{i < j} b_{ij}^2.
\end{aligned}
\end{equation}
As the estimator is unbiased, its Mean Squared Error (MSE) is equal to its variance. Following the bounds established in the proof of Theorem 6, where $\sum_{i < j} b_{ij}^2 \le nd_{\max}^3$, we conclude:
\[
    \text{MSE}\left( \frac{1}{6} \sum_{u=1}^n t_u \right) \le \frac{1}{9} \sigma^2 nd_{\max}^3 = O(nd_{\max}^3).
\]
\end{proof}

\subsection{Proof of \autoref{thm:5}}

\begin{proof}
Following the notation in \autoref{proof:thm:1}, let $\hat{A} = A + X$. For each node $u$, we define the local observation $t_u$ as:
\begin{equation*}
    t_u = \sum_{i} a_{ui}\hat{b}_{iu} = \sum_{i,j} a_{ui}(a_{ij} + x_{ij})(a_{ju} + x_{ju}),
\end{equation*}
where $\hat{b}_{iu}$ represents the noisy count of paths of length 2 between $i$ and $u$. Expanding the global sum $\sum_{u=1}^n t_u$, we obtain:
\begin{equation*}
\begin{aligned}
    \sum_{u=1}^n t_u &= \sum_{u,i,j} a_{ui}(a_{ij} + x_{ij})(a_{ju} + x_{ju}) \\
    &= \sum_{u,i,j} (a_{ui}a_{ij}a_{ju} + a_{ui}x_{ij}a_{ju} + a_{ui}a_{ij}x_{ju} + a_{ui}x_{ij}x_{ju}).
\end{aligned}
\end{equation*}
Utilizing the symmetry of matrices $A$ and $X$, and the fact that $\sum_{u,i,j} a_{ui}a_{ij}a_{ju} = \text{tr}(A^3) = 6f^{\triangle}(G)$, the expression simplifies to:
\begin{equation*}
\begin{aligned}
    \sum_{u=1}^n t_u = 6f^{\triangle}(G) &+ 2\sum_{i,j,k} a_{ij}a_{jk}x_{ki} + \sum_{i,j,k} a_{ij}x_{jk}x_{ki}.
\end{aligned}
\end{equation*}
From the properties established in \autoref{proof:thm:1}, the expectations of the noise terms are zero:
\[
    \mathbb{E}\left[ \sum_{i,j,k} a_{ij}a_{jk}x_{ki} \right] = 0, \quad \mathbb{E}\left[ \sum_{i,j,k} a_{ij}x_{jk}x_{ki} \right] = 0.
\]
Thus, the estimator is unbiased: $\mathbb{E}[ \frac{1}{6}\sum_{u=1}^n t_u ] = f^{\triangle}(G)$. 

The variance of the estimator is derived from the individual variance components established in \autoref{eq:var_first} and \autoref{eq:var_second}. Specifically:
\begin{equation*}
\label{eq:var_t_u_final}
\begin{aligned}
    \mathbb{V}\left( \frac{1}{6} \sum_{u=1}^n t_u \right) &= \frac{4}{36} \mathbb{V}\left( \sum_{i,j,k} a_{ij}a_{jk}x_{ki} \right) + \frac{1}{36} \mathbb{V}\left( \sum_{i,j,k} a_{ij}x_{jk}x_{ki} \right) \\
    &= \frac{4}{9} \sigma^2 \sum_{i < j} b_{ij}^2 + \frac{1}{9} \sigma^4 (n-2)|E|.
\end{aligned}
\end{equation*}
In \autoref{proof:thm:4}, we discussed $\sum_{i < j} b_{ij}^2 \le nd_{\max}^3$ and $|E| \le \frac{1}{2}nd_{\max}$, then MSE is bounded by:
\[
    \text{MSE}\left( \frac{1}{6} \sum_{u=1}^n t_u \right) \le \frac{4}{9} \sigma^2 nd_{\max}^3 + \frac{1}{18} \sigma^4 n^2 d_{\max},
\]
which yields the asymptotic complexity $O(nd_{\max}^3 + n^2d_{\max})$.
\end{proof}

\subsection{Proof of \autoref{thm:6}}

\begin{proof}
Following the notation in \autoref{proof:thm:1}, let $\hat{A} = A + X$. For each node $u$, we define the local estimator $t_u$ based on the noisy two-step path counts $\hat{b}_{ij}$:
\begin{equation*}
\begin{aligned}
    q_u &= \sum_{\left( i,j \right):a_{ui}=a_{ju}=1} (\hat{b}_{ij} - 1) \\
    &= \sum_{i,j} a_{ui}a_{uj} \left( \sum_{k} (a_{ik} + x_{ik})(a_{kj} + x_{kj}) - 1 \right) \\
    &= \sum_{i,j,k} a_{ui}a_{uj}a_{ik}a_{kj} - \sum_{i,j} a_{ui}a_{uj} \\
    &\quad + 2\sum_{i,j,k} a_{ui}a_{ij}a_{jk}x_{ku} + \sum_{i,j,k} a_{ui}a_{ij}x_{jk}x_{ku}.
\end{aligned}
\end{equation*}
The term $\sum_{i,j,k} a_{ui}a_{uj}a_{ik}a_{kj}$ represents the number of 4-step walks starting and ending at node $u$. The term $\sum_{i,j} a_{ui}a_{uj}$ accounts for 4-step walks that do not form a distinct quadrangle, specifically: (1) $u \to i \to u \to i \to u$ (where $i=j$), and (2) $u \to i \to u \to j \to u$ (where $i \neq j$). Subtracting these degenerate paths leaves walks that traverse a quadrangle. Since each quadrangle is traversed in both directions, we have:
\begin{equation*}
    \sum_{i,j,k} a_{ui}a_{uj}a_{ik}a_{kj} - \sum_{i,j} a_{ui}a_{uj} = 2f^\Box_u,
\end{equation*}
where $f^\Box_u$ is the number of quadrangles containing node $u$. Summing over all $u$ and noting that $\sum_{u=1}^n f^\Box_u = 4f^\Box(G)$, we obtain the unbiasedness:
\[
    \mathbb{E}\left[ \frac{1}{8} \sum_{u=1}^n q_u \right] = \frac{1}{8} \cdot 4 \cdot 2 f^\Box(G) = f^\Box(G).
\]

Next, we analyze the variance of the estimator. Firstly, by \autoref{lem:2}, the variance of the sum is the sum of the variances of the first-order and second-order noise components:
\begin{equation*}
    \mathbb{V}\left[ \sum_{u=1}^n q_u \right] = 4\mathbb{V}\left[ \sum_{u,i,j,k} a_{ui}a_{ij}a_{jk}x_{ku} \right] + \mathbb{V}\left[ \sum_{u,i,j,k} a_{ui}a_{ij}x_{jk}x_{ku} \right]
\end{equation*}

For the first component, let $c_{ij}$ denote the number of 3-step paths between $i$ and $j$, i.e., $c_{ij} = \sum_{k,l} a_{ik}a_{kl}a_{lj}$. Due to symmetry, the term simplifies to $2\sum_{i<j} c_{ij}x_{ij}$. Its variance is:
\begin{equation}
\label{eq:var_quad_first}
    4\mathbb{V}\left[ \sum_{i<j} 2c_{ij}x_{ij} \right] = 16\sigma^2 \sum_{i<j} c_{ij}^2.
\end{equation}

For the second component, the term can be rewritten as $\sum_{i,j} 2b_{ij} \sum_{k} x_{ik}x_{kj}$, where $b_{ij}$ is the number of 2-step paths. Its variance contribution is:
\begin{equation}
\label{eq:var_quad_second}
    \mathbb{V}\left[ \sum_{i<j} 2b_{ij} \sum_{k} x_{ik}x_{kj} \right] = 4(n-2)\sigma^4 \sum_{i<j} b_{ij}^2.
\end{equation}

Combining these results, the MSE of the global estimator is:
\begin{equation}
\label{eq:mse_quad_final}
    \text{MSE} = \frac{1}{4}\sigma^2 \sum_{i<j} c_{ij}^2 + \frac{1}{16}(n-2)\sigma^4 \sum_{i<j} b_{ij}^2.
\end{equation}

In a graph with maximum degree $d_{\max}$, we have $\sum_{i<j} c_{ij} \le nd_{\max}^3$ and each $c_{ij} \le d_{\max}^2$. To maximize $\sum c_{ij}^2$, we let $c_{ij} = d_{\max}^2$, which occurs for at most $nd_{\max}$ pairs. Thus, $\sum c_{ij}^2 \le nd_{\max}^5$. Similarly, $\sum b_{ij}^2 \le nd_{\max}^3$. Substituting these into \autoref{eq:mse_quad_final}:
\[
    \text{MSE} \le \frac{1}{4}\sigma^2 nd_{\max}^5 + \frac{1}{16}\sigma^4 n^2 d_{\max}^3 = O(nd_{\max}^5 + n^2d_{\max}^3).
\]
\end{proof}

\subsection{Proof of \autoref{thm:dprd}}

\begin{proof}
Fix an arbitrary ordered pair of neighboring datasets \(D,D'\in\mathcal{X}^n\) and an arbitrary measurable event \(A\subseteq\Xi\times\mathcal{Z}\). For every \(\eta\in\Xi\), define the \(\eta\)-section of \(A\) by
\[
A_\eta = \{z\in\mathcal{Z} : (\eta,z)\in A\},
\]
and set
\[
\begin{aligned}
&a(\eta)=\Pr_{\mathcal{R}_2}\bigl[\mathcal{R}_2(f(D,\eta))\in A_\eta\bigr],\\
&b(\eta)=\Pr_{\mathcal{R}_2}\bigl[\mathcal{R}_2(f(D',\eta))\in A_\eta\bigr]. 
\end{aligned}
\]
The probabilities are taken solely over the randomness of \(\mathcal{R}_2\); hence \(a,b:\Xi\to[0,1]\) are measurable functions. By the law of total probability,
\[
\begin{aligned}
&\Pr[\mathcal{M}(D)\in A] = \mathbb{E}_{\xi_D\sim P_D}\bigl[a(\xi_D)\bigr],\\
&\Pr[\mathcal{M}(D')\in A] = \mathbb{E}_{\xi_{D'}\sim P_{D'}}\bigl[b(\xi_{D'})\bigr].
\end{aligned}
\]

We first extract a positive‑part bound from the DPRD property of \(\mathcal{R}_2\).
Because \(\mathcal{R}_2\) provides \((\varepsilon_2,\delta_2)\)-DPRD with respect to \(P_D\), we have, for every measurable family of output events \(\{S_\eta\subseteq\mathcal{Z} : \eta\in\Xi\}\),
\[
\begin{aligned}
&\mathbb{E}_{\xi\sim P_D}\!\Pr_{\mathcal{R}_2}\bigl[\mathcal{R}_2(f(D,\xi))\in S_\xi\bigr]\\
\leq  & e^{\varepsilon_2}\,\mathbb{E}_{\xi\sim P_D}\!\Pr_{\mathcal{R}_2}\bigl[\mathcal{R}_2(f(D',\xi))\in S_\xi\bigr] + \delta_2.
\end{aligned}
\]
Apply this inequality to the family
\[
S_\eta =
\begin{cases}
A_\eta, & \text{if } a(\eta) > e^{\varepsilon_2}b(\eta),\\
\varnothing, & \text{otherwise}.
\end{cases}
\]
The left‑hand side becomes \(\mathbb{E}_{\xi\sim P_D}\bigl[a(\xi)\,\mathbf{1}_{\{a(\xi)>e^{\varepsilon_2}b(\xi)\}}\bigr]\), while the right‑hand side equals \(e^{\varepsilon_2}\,\mathbb{E}_{\xi\sim P_D}\bigl[b(\xi)\,\mathbf{1}_{\{a(\xi)>e^{\varepsilon_2}b(\xi)\}}\bigr]+\delta_2\). Taking the difference yields
\begin{equation}\label{eq:pos-part}
\mathbb{E}_{\xi\sim P_D}\Bigl[\bigl(a(\xi)-e^{\varepsilon_2}b(\xi)\bigr)_+\Bigr] \leq \delta_2,
\end{equation}
where \((x)_+ = \max\{0,x\}\).

Now define the truncated function
\[
\widetilde{a}(\eta) = a(\eta) - \bigl(a(\eta)-e^{\varepsilon_2}b(\eta)\bigr)_+ = \min\!\bigl\{a(\eta),\, e^{\varepsilon_2}b(\eta)\bigr\}.
\]
Clearly \(0\leq \widetilde{a}(\eta)\leq a(\eta)\leq 1\) and, by construction,
\begin{equation}\label{eq:tildea-bound}
\widetilde{a}(\eta) \leq e^{\varepsilon_2}b(\eta) \quad \text{for all }\eta\in\Xi.
\end{equation}
From~\eqref{eq:pos-part} we obtain the basic estimate
\begin{equation}\label{eq:a-tilde-delta}
\mathbb{E}_{P_D}[a] = \mathbb{E}_{P_D}[\widetilde{a}] + \mathbb{E}_{P_D}\Bigl[\bigl(a-e^{\varepsilon_2}b\bigr)_+\Bigr]
\leq \mathbb{E}_{P_D}[\widetilde{a}] + \delta_2.
\end{equation}

The first‑stage mechanism \(\mathcal{R}_1\) satisfies \((\varepsilon_1,\delta_1)\)-DP. In its expectation form this means that for every measurable function \(h:\Xi\to[0,1]\),
\[
\mathbb{E}_{\xi_D\sim P_D}\bigl[h(\xi_D)\bigr] \leq e^{\varepsilon_1}\,\mathbb{E}_{\xi_{D'}\sim P_{D'}}\bigl[h(\xi_{D'})\bigr] + \delta_1.
\]
Applying this inequality to \(h=\widetilde{a}\) (which indeed maps \(\Xi\) into \([0,1]\)) gives
\[
\mathbb{E}_{P_D}[\widetilde{a}] \leq e^{\varepsilon_1}\,\mathbb{E}_{P_{D'}}[\widetilde{a}] + \delta_1.
\]
Using~\eqref{eq:tildea-bound} to replace \(\widetilde{a}\) by \(e^{\varepsilon_2}b\) inside the expectation on \(P_{D'}\), we obtain
\[
\mathbb{E}_{P_D}[\widetilde{a}] \leq e^{\varepsilon_1+\varepsilon_2}\,\mathbb{E}_{P_{D'}}[b] + \delta_1.
\]
Finally, combine this with~\eqref{eq:a-tilde-delta}:
\[
\begin{aligned}
&\Pr[\mathcal{M}(D)\in A] = \mathbb{E}_{P_D}[a] \\
\leq &e^{\varepsilon_1+\varepsilon_2}\,\mathbb{E}_{P_{D'}}[b] + \delta_1 + \delta_2 \\
= &e^{\varepsilon_1+\varepsilon_2}\,\Pr[\mathcal{M}(D')\in A] + \delta_1 + \delta_2.
\end{aligned}
\]
Since the neighboring datasets \(D,D'\) and the event \(A\subseteq\Xi\times\mathcal{Z}\) were arbitrary, the mechanism \(\mathcal{M}\) satisfies \((\varepsilon_1+\varepsilon_2,\delta_1+\delta_2)\)-differential privacy.
\end{proof}

\subsection{Proof of \autoref{pro:clip}}

\begin{proof}
Let $Y = X - \mu$. Since $X$ is symmetric about $\mu$, $Y$ is symmetric about $0$. For $\beta \in [0,0.5]$, let $y_\beta \ge 0$ be such that
\[
q_\beta^+ = \mu + y_\beta,\qquad q_\beta^- = \mu - y_\beta .
\]
Assuming $\mu > 0$ (otherwise apply the argument to $-X$), we have
\[
\kappa_\beta = \max\{|q_\beta^+|, |q_\beta^-|\} = \mu + y_\beta .
\]
Let $X_c = \operatorname{clip}(X,\kappa_\beta)$. The absolute bias is
\[
\Delta = \mathbb E[X - X_c].
\]
Since $X_c = X$ on $[-\kappa_\beta,\kappa_\beta]$, while $X_c = \kappa_\beta$ for $X > \kappa_\beta$ and $X_c = -\kappa_\beta$ for $X < -\kappa_\beta$, we obtain
\[
\begin{aligned}
\Delta
&= \mathbb E\bigl[(X-\kappa_\beta)\mathbf 1_{\{X>\kappa_\beta\}}\bigr]
 + \mathbb E\bigl[(X+\kappa_\beta)\mathbf 1_{\{X<-\kappa_\beta\}}\bigr] \\
&= \mathbb E\bigl[(Y-y_\beta)\mathbf 1_{\{Y>y_\beta\}}\bigr]
 + \mathbb E\bigl[(Y+2\mu+y_\beta)\mathbf 1_{\{Y<-2\mu-y_\beta\}}\bigr].
\end{aligned}
\]
By symmetry of $Y$ and the change of variables $Z=-Y$, the second term equals
\[
-\mathbb E\bigl[(Y-(2\mu+y_\beta))\mathbf 1_{\{Y>2\mu+y_\beta\}}\bigr].
\]
Hence, using the identity
\[
\mathbb E[(Y-a)\mathbf 1_{\{Y>a\}}] = \int_a^\infty \Pr(Y>t)\,dt,
\]
we get
\[
\Delta = \int_{y_\beta}^{2\mu+y_\beta} \Pr(Y>t)\,dt .
\]
Since $\Pr(Y>y_\beta)=\beta$ and the survival function $t\mapsto \Pr(Y>t)$ is nonincreasing, for every $t\in[y_\beta,2\mu+y_\beta]$ we have
\[
0 \le \Pr(Y>t) \le \beta .
\]
Therefore,
\[
0 \le \Delta \le 2\mu\beta .
\]
Dividing by $\mu>0$ yields
\[
0 \le \frac{\mathbb E[X]-\mathbb E[\operatorname{clip}(X,\kappa_\beta)]}{\mathbb E[X]} \le 2\beta .
\]
\end{proof}

\subsection{Proof of \autoref{thm:2RPureDP}}

\begin{proof}

Clipping imposes an upper bound \(\kappa\) or \(\kappa_u\) on each element of \(\hat{A}\), \(\hat{B}\), or \(\hat{\mathbb{b}}_u\), which removes randomness. Therefore, the remaining proof only needs to consider whether \(2\kappa \tilde{d}_u\) or \(\kappa_u\) covers the sensitivity in the worst case.

For TriTR and QuaTR: For user \(u\), the true degree after graph projection satisfies \(d_u \leq \tilde{d}_u\). Adding or removing one neighbor changes \(\text{sum}_u\) by at most the sum of elements in one row and one column (see \autoref{fig:4}), involving at most \(2 d_u\) elements of \(\hat{A}\) or \(\hat{B}\). Since their absolute values are clipped within \([-\kappa, \kappa]\), the change is at most \(2\kappa \tilde{d}_u\). Therefore, adding \(\text{Lap}(2\kappa \tilde{d}_u/\epsilon_2)\) noise suffices to preserve \(\epsilon_2\)-edge LDP in the second round. Finally, applying \autoref{pro:2} and \autoref{pro:1} gives the claimed result.

For TriMTR: For user \(u\), the change in \(\text{sum}_u\) from adding or removing one neighbor is a single element of \(\hat{\mathbb{b}}_u\), whose magnitude after clipping is bounded by \([-\kappa_u, \kappa_u]\). The same reasoning yields the final conclusion.

\end{proof}

\subsection{Proof of \autoref{thm:epsilon_delta_DP}}
\label{sec:E}

\begin{proof}

To facilitate subsequent proof, we first prove \autoref{lem:dprd_sufficient}:

\begin{lemma}\label{lem:dprd_sufficient}
Let \(\mathcal{M}_{P_{\Xi}}\) be defined as in ~\autoref{def:DPRD}. Suppose there exists a measurable subset \(E \subseteq \Xi\) such that
\[
P_{\Xi}(E) \ge 1 - \delta,
\]
and for any neighboring datasets \(D, D' \in \mathcal{X}^n\), any measurable family of output events \(\{S_\eta \subseteq \mathcal{Z} : \eta \in \Xi\}\), and every \(\xi \in E\), the conditional inequality
\begin{equation}
\begin{aligned}
\Pr_{\mathcal{R}}\left[\mathcal{R}(f(D,\xi)) \in S_\xi\right]
\le
e^{\epsilon}\,
\Pr_{\mathcal{R}}\left[\mathcal{R}(f(D',\xi)) \in S_\xi\right]
\end{aligned}
\label{equ:114514}
\end{equation}
holds. Then \(\mathcal{M}_{P_{\Xi}}\) satisfies \((\epsilon,\delta)\)-differential privacy on randomized data with respect to \(P_{\Xi}\).
\end{lemma}

\begin{proof}
Fix arbitrary neighboring datasets \(D, D' \in \mathcal{X}^n\) and an arbitrary measurable family \(\{S_\eta \subseteq \mathcal{Z} : \eta \in \Xi\}\). For each \(\xi \in \Xi\), define the conditional probability
\[
p_D(\xi) := \Pr_{\mathcal{R}}\left[\mathcal{R}(f(D,\xi)) \in S_\xi\right],
\]
and similarly define \(p_{D'}(\xi)\). The joint probability over \(\xi \sim P_{\Xi}\) and the mechanism \(\mathcal{R}\) is given by \(\mathbb{E}_{\xi \sim P_{\Xi}}[p_D(\xi)]\). Decomposing the expectation over \(E\) and its complement \(E^c\), we obtain
\[
\begin{aligned}
\mathbb{E}_{\xi}[p_D(\xi)]
&= \int_E p_D(\xi) \, dP_{\Xi}(\xi) + \int_{E^c} p_D(\xi) \, dP_{\Xi}(\xi) \\
&\le \int_E e^{\epsilon} p_{D'}(\xi) \, dP_{\Xi}(\xi) + \int_{E^c} 1 \, dP_{\Xi}(\xi) \\
&\le e^{\epsilon} \int_{\Xi} p_{D'}(\xi) \, dP_{\Xi}(\xi) + P_{\Xi}(E^c) \\
&\le e^{\epsilon} \, \mathbb{E}_{\xi}[p_{D'}(\xi)] + \delta,
\end{aligned}
\]
where the first inequality follows from \eqref{equ:114514} on \(E\) and the trivial bound \(p_D(\xi) \le 1\) on \(E^c\), and the last inequality uses the assumption \(P_{\Xi}(E^c) \le \delta\). Substituting the definition of \(p_D\) and \(p_{D'}\) yields
\[
\Pr_{\xi \sim P_{\Xi}, \mathcal{R}}
\!\left[
\mathcal{R}(f(D,\xi)) \in S_{\xi}
\right]
\leq
e^{\epsilon}
\Pr_{\xi \sim P_{\Xi}, \mathcal{R}}
\!\left[
\mathcal{R}(f(D',\xi)) \in S_{\xi}
\right]
+\delta.
\]
Since \(D, D'\), and the family \(\{S_\eta\}\) were chosen arbitrarily, this is precisely the condition in \autoref{def:DPRD}. Hence \(\mathcal{M}_{P_{\Xi}}\) satisfies \((\epsilon, \delta)\)-DPRD with respect to \(P_{\Xi}\).
\end{proof}

We set \(K=\{\hat{A}: \| \sum_{i \in \mathbf{adj}_u} \hat{a}_{ij}\| \leq \kappa_u, \ \text{for any }   j \in \mathbf{adj}_u \}\). Then we only need to prove that \(P(K^c) \leq \delta\). Further, it suffices to show that for each fixed \(j \in \mathbf{adj}_u\),
\[
P \left( \left| \sum_{i \in \mathbf{adj}_u} \hat{a}_{ij} \right| > \kappa_u  \right) \leq \frac{\delta}{d_u}.
\]
Moreover, it is enough to prove the upper-tail bound
\[
P \left( \sum_{i \in \mathbf{adj}_u} \hat{a}_{ij} > \kappa_u\right) \leq \frac{\delta}{2d_u},
\]
and the lower-tail bound
\[
P \left( \sum_{i \in \mathbf{adj}_u} \hat{a}_{ij} < -\kappa_u \right) \leq \frac{\delta}{2d_u}
\]
follows by \(a_{ij} \geq 0\).

We now state the Bernstein inequality that will be used.

\begin{lemma}[Sub-exponential Bernstein inequality]\label{lem:bernstein_subexp}
Let \(Y_1,\dots,Y_n\) be independent centered random variables. Suppose there exist constants \(\sigma^2>0\) and \(\nu>0\) such that for all \(|\lambda|<\frac{1}{\nu}\),
\[
\mathbb{E}\!\left[ e^{\lambda Y_i} \right]
\le
\exp\!\left( \frac{\sigma^2 \lambda^2}{2(1-\nu|\lambda|)} \right).
\]
Then for all \(t\ge 0\),
\[
\Pr\!\left( \sum_{i=1}^n Y_i \ge t \right)
\le
\exp\!\left( -\frac{t^2}{2n\sigma^2 + 2\nu t} \right).
\]
\end{lemma}

For the Laplace mechanism, if \(\hat{a}_{ij} \sim \mathrm{Lap}(a_{ij}, 1/\epsilon_1)\), then \(Y_i=\hat{a}_{ij}-a_{ij}\sim \mathrm{Lap}(0,1/\epsilon_1)\). Its variance is \(\sigma^2 = 2/\epsilon_1^2\), and it satisfies the condition of \autoref{lem:bernstein_subexp} with \(\nu = 1/\epsilon_1\).

For the randomized response mechanism, the variable \(\hat{a}_{ij}\) takes two values, and its centered version \(Y_i = \hat{a}_{ij} - a_{ij}\) is bounded in absolute value by
\[
M = \frac{e^{\epsilon_1}}{e^{\epsilon_1}-1}.
\]
A standard bound on the moment generating function of a centered bounded variable gives
\[
\mathbb{E}\!\left[ e^{\lambda Y_i} \right]
\le
\exp\!\left( \frac{\sigma^2 \lambda^2}{2(1-\frac{M}{3}|\lambda|)} \right),
\qquad |\lambda| < \frac{3}{M},
\]
where \(\sigma^2 = \mathbb{V}(\hat{a}_{ij})\). Hence it also satisfies \autoref{lem:bernstein_subexp} with \(\nu = M/3 = \frac{e^{\epsilon_1}}{3(e^{\epsilon_1}-1)}\).

We now prove the upper-tail bound. Fix an arbitrary \(j \in \mathbf{adj}_u\), and let \(d = d_u\) and \(\mathcal{I} = \mathbf{adj}_u\). For each \(i \in \mathcal{I}\), define
\[
Y_i = \hat{a}_{ij} - a_{ij}.
\]
Then \(Y_i\) are independent and centered, and they satisfy the Bernstein condition with common parameters \(\sigma^2\) and \(\nu\). Moreover, since \(a_{ij}\in\{0,1\}\) and \(|\mathcal{I}|=d\),
\[
\sum_{i\in\mathcal{I}} a_{ij} \le d.
\]
Thus,
\[
\sum_{i\in\mathcal{I}} \hat{a}_{ij}
= \sum_{i\in\mathcal{I}} a_{ij} + \sum_{i\in\mathcal{I}} Y_i
\le d + \sum_{i\in\mathcal{I}} Y_i.
\]
Therefore,
\[
\Pr\!\left( \sum_{i\in\mathcal{I}} \hat{a}_{ij} > d + t \right)
\le
\Pr\!\left( \sum_{i\in\mathcal{I}} Y_i > t \right)
\le
\exp\!\left( -\frac{t^2}{2d\sigma^2 + 2\nu t} \right).
\]
Let \(L = \ln(2d/\delta)\). Choose
\[
t = \nu L + \sqrt{ \nu^2 L^2 + 2d\sigma^2 L }.
\]
Then \(t\) solves the quadratic equation
\[
t^2 = L(2d\sigma^2 + 2\nu t),
\]
and hence
\[
\exp\!\left( -\frac{t^2}{2d\sigma^2 + 2\nu t} \right)
= e^{-L}
= \frac{\delta}{2d}.
\]
It follows that
\[
\Pr\!\left( \sum_{i\in\mathcal{I}} \hat{a}_{ij} > \kappa_u \right)
\le
\frac{\delta}{2d},
\qquad
\kappa_u = d + \nu L + \sqrt{ \nu^2 L^2 + 2d\sigma^2 L }.
\]
The lower-tail bound is obtained by applying the same argument to \(-Y_i\), which satisfies the same Bernstein condition. Since
\[
\sum_{i\in\mathcal{I}} \hat{a}_{ij}
= \sum_{i\in\mathcal{I}} a_{ij} + \sum_{i\in\mathcal{I}} Y_i
\ge \sum_{i\in\mathcal{I}} Y_i,
\]
the event \(\sum_{i\in\mathcal{I}} \hat{a}_{ij} < -\kappa_u\) implies \(\sum_{i\in\mathcal{I}} Y_i < -t\) with \(t = \kappa_u - d\). Thus,
\[
\Pr\!\left( \sum_{i\in\mathcal{I}} \hat{a}_{ij} < -\kappa_u \right)
\le
\Pr\!\left( \sum_{i\in\mathcal{I}} Y_i < -t \right)
\le
\frac{\delta}{2d}.
\]
Combining the two tails, we obtain for every fixed \(j\),
\[
\Pr\!\left( \left| \sum_{i \in \mathbf{adj}_u} \hat{a}_{ij} \right| > \kappa_u \right)
\le
\frac{\delta}{d}.
\]
Taking a union bound over all \(j \in \mathbf{adj}_u\) gives
\[
\Pr(K^c)
\le
d_u \cdot \frac{\delta}{d_u}
=
\delta.
\]
Thus \(P(K) \ge 1-\delta\).

In the algorithm, the threshold \(\kappa_u\) is computed using the noisy degree \(\tilde{d}_u\). Since \(\tilde{d}_u \ge d_u\) by construction, and \(\kappa_u(\tilde{d}_u) \ge \kappa_u(d_u)\), the same coverage guarantee remains valid.

Finally, the proof is completed by applying \autoref{lem:dprd_sufficient}, \autoref{thm:dprd}, \autoref{pro:2}, and \autoref{pro:1} in this order.

\end{proof}

\subsection{Proof of \autoref{pro:MSE}}

\begin{proof}
In the proofs of \autoref{thm:4}--\ref{thm:5}, we have given the full expression and upper bounds for the one-round noise variance. Their upper bounds are, respectively, \(O(nd_{\max}^3)\), \(O(nd_{\max}^3 + n^2d_{\max})\), and \(O(nd_{\max}^5 + n^2d_{\max}^3)\). Next, we only need to give the upper bounds for their two-round noise.

Let \(\mathbb{V}_2(\mathrm{TriTR})\), \(\mathbb{V}_2(\mathrm{TriMTR})\), and \(\mathbb{V}_2(\mathrm{QuaTR})\) denote the variances of the noise added in their second round, respectively. Then:
\begin{align*}
\MoveEqLeft[2] \mathbb{V}_2(\mathrm{TriTR}) \\
&= \sum_{i=1}^n 2\left(\frac{2\kappa\tilde{d}_i}{\epsilon_2} \right)^2 \\
&= O\left( \sum_{i=1}^n d_i^2 \right) \le O(nd_{\max}^2), \\[1ex]
\MoveEqLeft[2] \mathbb{V}_2(\mathrm{TriMTR}) \\
&= \sum_{i=1}^n 2\left(\frac{\kappa_u}{\epsilon_2} \right)^2 \\
&= O\left( \sum_{i=1}^n (d_i+\sqrt{n})^2 \right) \le O(nd_{\max}^2+n^2), \\[1ex]
\MoveEqLeft[2] \mathbb{V}_2(\mathrm{QuaTR}) \\
&= \sum_{i=1}^n 2\left(\frac{2\kappa\tilde{d}_i}{\epsilon_2} \right)^2 \\
&= O\left( \sum_{i=1}^n \bigl((d_i+\sqrt{n})d_i\bigr)^2 \right) \le O(nd_{\max}^4 + n^2d_{\max}^2).
\end{align*}

Therefore, the MSE upper bounds for TriTR, TriMTR, and QuaTR are respectively \(O(nd_{\max}^3)\), \(O(nd_{\max}^3 + n^2d_{\max})\), and \(O(nd_{\max}^5 + n^2d_{\max}^3)\). Since \(\text{TriTR}^*\) is a reduced-noise version of TriTR, its MSE upper bound is also \(O(nd_{\max}^3)\).

\end{proof}

\subsection{Proof of \autoref{thm:TriRE}}

\begin{proof}
Let $C_g$ denote the global clustering coefficient. Based on the definition $C_g = \frac{3 \times f^\triangle(G)}{\#\text{2-stars}}$, where $\#\text{2-stars} = \sum_{i=1}^n \binom{d_i}{2}$, the true triangle count satisfies:
\begin{equation}
f^\triangle(G) = \frac{C_g}{3} \sum_{i=1}^n \frac{d_i(d_i-1)}{2} = \Theta\left( C_g \sum_{i=1}^n d_i^2 \right).
\label{eq:true_tri}    
\end{equation}

The relative error is bounded using Jensen's inequality as:
\[
    \mathrm{RE} \triangleq \mathbb{E}\left[\frac{|\hat f^\triangle(G)-f^\triangle(G)|}{f^\triangle(G)}\right] \le \frac{\sqrt{\mathrm{MSE}}}{f^\triangle(G)}.
\]
Substituting \autoref{eq:true_tri} into the above inequality, we obtain:
\[
    \mathrm{RE} \le O\left( \frac{1}{C_g} \cdot \frac{\sqrt{\mathrm{MSE}}}{\sum_{i=1}^n d_i^2} \right).
\]

\smallskip
\noindent\textbf{TriTR:}
\begin{align*}
    &\mathrm{RE}\\
    \leq & O\left( \frac{1}{C_g} \sqrt{\frac{\sum_{i < j} b_{ij}^2 + \sum_{i=1}^n d_i^2 }{(\sum_{i=1}^n d_i^2)^2}} \right) \\
    \le & O\left( \frac{1}{C_g} \sqrt{\frac{\sum_{i < j} d_i^2 + \sum_{i=1}^n d_i^2 }{(\sum_{i=1}^n d_i^2)^2}} \right) \\
    \le & O\left( \frac{1}{C_g} \sqrt{\frac{n\sum_{i =1}^n d_i^2 }{(\sum_{i=1}^n d_i^2)^2}} \right) \\
    = & O\left( \frac{1}{C_g} \sqrt{\frac{n}{\sum_{i=1}^n d_i^2}} \right) \\
    \leq & O\left( \frac{1}{C_g} \sqrt{\frac{n}{n d_{avg}^2}} \right) \\
    = & O\left( \frac{1}{C_gd_{avg}}\right)
\end{align*}

\smallskip
\noindent\textbf{TriMTR:}
\begin{align*}
    &\mathrm{RE}\\
    \leq & O\left( \frac{1}{C_g} \sqrt{\frac{\sum_{i < j} b_{ij}^2 + n^2 d_{avg} + \sum_{i=1}^n (d_i+\sqrt{n})^2 }{(\sum_{i=1}^n d_i^2)^2}} \right) \\
    \le & O\left( \frac{1}{C_g} \sqrt{\frac{n\sum_{i =1}^n d_i^2 +  n^2 d_{avg}}{(\sum_{i=1}^n d_i^2)^2}} \right) \\
    = & O\left( \frac{1}{C_g} \sqrt{\frac{n\sum_{i =1}^n d_i^2 }{(\sum_{i=1}^n d_i^2)^2}  + \frac{n^2 d_{avg}}{(\sum_{i=1}^n d_i^2)^2} } \right) \\
    = & O\left( \frac{1}{C_g} \sqrt{\frac{n}{\sum_{i=1}^n d_i^2}  + \frac{n^2 d_{avg}}{(\sum_{i=1}^n d_i^2)^2} } \right) \\
    \leq & O\left( \frac{1}{C_g} \sqrt{\frac{n}{nd_{avg}^2}  + \frac{n^2 d_{avg}}{n^2d_{avg}^4} } \right) \\
    = & O\left( \frac{1}{C_g} \sqrt{\frac{1}{d_{avg}^2}  + \frac{1}{d_{avg}^3} } \right) \\
    = & O\left( \frac{1}{C_gd_{avg}}\right)
\end{align*}
This completes the proof.
\end{proof}

\end{document}